\documentclass[fleqn,usenatbib]{mnras}

\usepackage{newtxtext,newtxmath}

\usepackage[T1]{fontenc}
\usepackage{ae,aecompl}

\usepackage{graphicx,subfigure}	
\usepackage{amsmath}	
\usepackage{textgreek} 
\usepackage{placeins} 
\usepackage{wasysym} 
\usepackage{hyperref} 
\usepackage{enumitem}     

\title{Low-redshift quasars in the SDSS Stripe 82 II: associated companion galaxies and signature of star formation}

\author[M.B.Stone]{
M. B. Stone,$^{1,2}$\thanks{E-mail: \href{mailto:mbston@utu.fi}{mbstone12@gmail.com}}
D. Bettoni,$^{3}$
R. Falomo,$^{3}$
J. K. Kotilainen,$^{2,1}$
K. Karhunen,$^{1}$
S. Paiano$^{4,5}$
\newauthor{ and R. Scarpa$^{6,7}$}
\\
$^{1}$Department of Physics and Astronomy, University of Turku, 20014 Turku, Finland\\
$^{2}$Finnish Centre for Astronomy with ESO (FINCA), University of Turku, 20014 Turku, Finland\\
$^{3}$INAF - Osservatorio Astronomico di Padova, Vicolo dell'Osservatorio 5, I-35122 Padova (PD), Italy\\
$^{4}$Osservatorio Astronomico di Roma, via Frascati 33, I-00040 Monteporzio Catone, Italy\\
$^{5}$IASF Milano, via Corti 12, I-20133 Milano, Italy\\
$^{6}$Instituto de Astrofisica de Canarias, C/O Via Lactea, s/n E38205, La Laguna (Tenerife), Spain\\
$^{7}$Universidad de La Laguna, Dpto. Astrofsica, s/n E-38206, La Laguna (Tenerife), Spain
}

\date{Accepted XXX. Received YYY; in original form ZZZ}

\pubyear{2020}

\begin{document}
\label{firstpage}
\pagerange{\pageref{firstpage}--\pageref{lastpage}}
\maketitle

\begin{abstract}

We present optical spectroscopy of the close companions of 22 low redshift (z\textless0.5) quasars (QSO) selected from a larger sample of QSO in the SDSS Stripe82 region for which both the host galaxy and the large scale environments have been investigated in our previous work.
The new observations extend the number of QSO studied in our previous paper on close companion galaxies of 12 quasars.
Our analysis here covers all 34 quasars from both this work and the previously published paper.
We find that half of them (15 QSO; $\sim$44\%) have at least one associated galaxy.
Many (12 galaxies; $\sim$67\%) of the associated companions exhibit [O\,\textsc{ii}] 3727 {\AA} emission line as signature of recent star formation.
The star formation rate (SFR) of these galaxies is modest (median SFR $\sim$ 4.3 M$_{\odot}$ yr$^{-1}$).
For 8 QSO we are also able to detect the starlight of the host galaxy from which 3 have a typical spectrum of a post-starburst galaxy.
Our results suggest that quasars do not have a strong influence on the star formation of their companion galaxies.

\end{abstract}

\begin{keywords}
galaxies: active -- galaxies: evolution -- galaxies: nuclei -- quasars: general.
\end{keywords}



\section{Introduction}

Active galactic nuclei (AGN) are a small fraction of galaxies whose supermassive black holes (SMBH) are actively accreting matter, a process which results in formidable energy outputs.
Strong clues of the importance of SMBHs in the formation and evolution of the host galaxy are given by the correlations discovered between the mass of the SMBH and the properties of the host galaxies, such as mass or luminosity \citep{Magorrian_1998,Ferrarese_2006}, velocity dispersion of host bulge \citep{Ferrarese_2000,Gebhardt_2000,McConnell_2013,Kormendy_2013} and dark matter halos \citep{Ferrarese_2002}.
Advanced cosmological simulations of the universe (eg. Illustris project; \citealt{Vogelsberger_2014}) suggest the importance of energetic feedback processes from accreting SMBHs \citep{Somerville_2015,Naab_2017}.
However, the mechanism(s) of how the nuclear activity is triggered and maintained represent still an open question.

One possible method of delivering fuel material for SMBH accretion are minor and major galactic mergers \citep{Toomre_1972,Byrd_1986,Sanders_1988,Kauffmann_2000,Hopkins_2006_merger,Glikman_2015,Villforth_2015,Weston_2017,Villforth_2017,Tao_2017,Martin_2018,Gordon_2019}.
Hydrodynamical simulations of galaxy-galaxy collisions illustrate the generation of star formation bursts and strong inflows which could rapidly feed the central SMBH \citep{DiMatteo_2005}.
On the other hand, a body of research shows secular processes to drive the fueling of SMBHs \citep{Bahcall_1996, Hopkins_2006_secular,Khalatyan_2008,Treister_2012, Smethurst_2019}.
It is therefore of interest to observe the close environments of QSO to investigate whether there are nearby galaxies that could participate in a merger and whether recent star formation as well as black hole growth and regulation are correlated with the interactions between galaxies.

On the global scale of supercluster-void network, quasars trace the Large Scale Structure, however, their location is constrained to the less dense peripheries of the clusters \citep{Sochting_2001,Lietzen_2009}. 
Cluster mergers could play a role in quasar formation \citep{Sochting_2002}, while global environmental properties can influence their evolution \citep{Kauffmann_2002,Lietzen_2009,Lietzen_2011,Shen_2017}.
Radio-selected AGN at redshifts z$\leq$1.2 are often found in more dense environments than inactive galaxies, however the star-formation activity of their host galaxies is not related to the large scale environment of the galaxy \citep{Magliocchetti_2018}.
The neighboring galaxies around giant radio galaxies (GRGs) and GRGs themselves show a larger proportion of intermediate age stellar populations, which could have resulted from past merger events or from cold stream activity in the galaxy groups \citep[e.g.][]{Kuzmicz_2019}.

In the vicinity of AGN, a physically associated `companion' galaxy is defined as a galaxy which may interact with the AGN gravitationally, but does not have to be gravitationally bound \citep{Peterson_1997}.
In the scenario where the nuclear activity is fueled by mergers, close companions may be galaxies which could merge with the quasar host and fuel the SMBH.
The first attempts to study the close environments around QSO suggested that the peculiar visible features in close proximity to quasars (tails, bridges, low surface brightness protrusions) could be due to past close encounters with other galaxies \citep{Toomre_1972,Stockton_1982}.
Some imaging studies (e.g. \citealt{Bennert_2008}) show fine structure abundance in QSO host galaxies, supporting the merger scenario.
A recent imaging study based on \textit{Hubble Space Telescope} data for $\sim$500 quasars at redshifts 0.3\textless z\textless3, compared faint, intermediate and bright companions of quasars and inactive galaxies \citep{Yue_2019}, finding that compared to inactive galaxies, quasars show a lower number of intermediate companions but similar numbers of faint and bright companions. 
Intermediate companion galaxies indicate major merger activity, and their deficit supports merger-triggered quasar evolution.

On the other hand alternative scenarios for explaining the triggering of the nuclear activity are also envisaged.
A number of studies looking at the morphological signatures of mergers in AGN host galaxies concluded that major mergers were not the dominant triggering process for the activation of the SMBH.
For a sample of 140 AGN host galaxies from the Cosmic Evolution Survey (COSMOS; \citealt{Scoville_2007}) in the redshift range 0.3\textless z\textless1.0 neither strong morphological distortions based on high-resolution images nor different distortion fractions were found compared to a matched sample of inactive galaxies \citep{Cisternas_2011}.
This supports the secular processes and minor mergers as the triggering mechanism for the SMBH activity at least for that cosmic epoch.
Subsequently, the morphological analysis of host galaxies based on high resolution imaging also suggests secular processes to be more frequent in SMBH triggering for high luminosity AGN (0.5\textless z\textless0.7; \citealt{Villforth_2017}), Iron Low-ionization Broad Absorption Line (FeLoBAL) quasars (0.6\textless z\textless1.1; \citealt{Villforth_2019}), and optically obscured (Type II) quasars (0.04\textless z\textless0.4; \citealt{Zhao_2019}).

In particular, the AGN fractions amongst galaxy pairs have also been assessed, indicating that if interactions result in ignition of AGN, then the merger stage should be completed before the SMBH activity \citep{Ellison_2008}.
Furthermore, the nuclear fuelling of low-excitation radio galaxies is not due to major interactions, but rather is activated by secular processes \citep{Ellison_2015}.
Moreover, \cite{Sol_2018} showed that both bar instability and interactions with other galaxies have the potential to fuel nuclear activity, however, the secular process is more efficient.

Further studies looked at the star formation in the neighborhoods of quasars since interactions may trigger star formation activity in the quasar and in the galaxies around the quasar.
Galaxies around active objects in the redshift range 0.1\textless z\textless0.2 have a higher star formation rate, than those around normal galaxies \citep{Coldwell_2003}.
Statistical analysis of the quasars from the Sloan Digital Sky Survey (SDSS) DR3 at z\textless0.2 confirmed that quasars do not reside in high-density regions and the environments of quasars, when compared to normal galaxies, show higher star formation activity, bluer colors and disc-type morphologies, however the triggering of the nuclear activity does not necessitate the presence of a close companion galaxy \citep{Coldwell_2006}.
The SDSS galaxy pairs show enhanced SFR in lower density environments, however no increase in star formation is shown for early-type galaxies even though the merger activity is widespread \citep{Ellison_2010}.

The [O\,\textsc{ii}] emission line properties were used to evaluate the SFR in quasars, showing that the SFR in quasars are typically very low, even though there is an abundance of gas \citep{Ho_2005}. 
Hence, some sort of suppression mechanism is in action to lower SFR in quasars.
\cite{Li_2008} state that while galaxy interactions are associated with enhanced star formation, their study failed to find any corresponding relation for the specific case of AGN activity.
Low redshift to intermediate redshift SDSS quasars have also been investigated for their [O\,\textsc{ii}] emission-line properties \citep{Kalfountzou_2012}, showing that the radio-loud quasar population has a higher SFR than radio-quiet quasar sample.

\defcitealias{Falomo_2014}{F14}

Interactions between QSO and physically associated companion galaxies may play a role in triggering and fuelling the nuclear activity and these processes may result in star formation in the QSO and/or in its companion galaxy, e.g. \citealt{DiMatteo_2005}.
To study this problem in a statistically sound manner we need a large number of observations accompanied with a similar analysis of ‘control’ fields of normal galaxies.
In our previous works we used the SDSS Stripe82 deep images to investigate the host galaxy and the environmental properties \citep{Karhunen_2014,Bettoni_2015} of a large homogeneous sample of low-redshift (z\textless0.5) quasars \citep[hereafter \citetalias{Falomo_2014}]{Falomo_2014}. 
We compared environments within a volume of 1 Mpc of this sample of quasars to a matching sample of inactive galaxies, concluding that no significant differences were found in galaxy number densities for active and inactive galaxies \citep{Karhunen_2014}.
The SDSS magnitude-based colors of hosts were further studied by \cite{Bettoni_2015} and the main conclusion was that the mean colours of the QSO host galaxy are very similar to those of a sample of inactive galaxies matched in terms of redshift and galaxy luminosity with the quasar sample.

\defcitealias{Bettoni_2017}{I}
The interaction of the AGN-host galaxy with its environment could lead to star formation events in companion galaxies.
We undertook a spectroscopic program of low-redshift quasars and their close companions with the aim to assess the recent star formation both in the QSO and in the companion galaxies. 
Our main goal is to measure the redshift of the galaxies in the immediate environments of the QSO to probe physical association and to search for signature of recent star format
on in the companion galaxies and when possible in the QSO host galaxies.
In \cite{Bettoni_2017}, hereafter Paper \citetalias{Bettoni_2017}, we presented results for 12 quasars at redshifts z\textless0.3.
Most (8 out of 12) of these quasars had an associated companion. The moderate SFR as derived from [O\,\textsc{ii}] luminosity, suggested that the QSO activity plays a modest role in the SF of nearby companions.
Here we extend the sample of QSO up to z$\sim$0.5 and present the results for 22 new QSO using observations of both the Gran Telescopio Canarias (GTC) and the Nordic Optical Telescope (NOT). 

This paper is organised as follows. 
The data sample is presented in Section 2. 
The analysis is presented in Section 3.
The results of this study
\footnote{
The new subsample is referred to as "this work", and we use the phrase "whole sample" when describing results from both the previous publication (Paper \citetalias{Bettoni_2017}) and this work.
}
and an overview including findings from Paper \citetalias{Bettoni_2017} data are presented in Section 4. 
Finally in Section 5 we summarise our main conclusions.
The results are obtained in the framework of the concordance cosmology, using $H_{0}$ = 70 km s$^{-1}$ Mpc$^{-1}$, $\Omega_{m}$ = 0.3, $\Omega_{\lambda}$ = 0.7.

\section{The sample}
The new QSO targets (0.2\textless z\textless0.5) were selected from the \citetalias{Falomo_2014} parent sample of 416 low-redshift QSO located in the Stripe 82 \citep{Annis_2014} region of the sky.
We have described for these QSO their local environments \citep{Karhunen_2014} and the host galaxy properties \citep{Bettoni_2015}.
Similarly to Paper \citetalias{Bettoni_2017}, each of the selected QSO had at least one companion galaxy candidate.

To select the candidate companion galaxies, we used the SDSS to retrieve their properties (coordinates, SDSS apparent magnitudes). 
We used visual inspection to inspect the QSO fields and confirm the candidate companion galaxies.
All candidate companion galaxies were selected to have an apparent magnitude brighter than SDSS m$_r$=22, as in Paper \citetalias{Bettoni_2017}, in order to be able to obtain adequate quality spectra.
The PD limit is set to 250 kpc from the QSO to the candidate companion galaxy.
In \citetalias{Falomo_2014} sample of quasars, there is an overdensity of galaxies within 200 kpc \citep{Karhunen_2014}.
Similar PD cutoff was used to define a close pair of galaxies in other studies \citep{Zitelli_2004,Focardi_2006,Behroozi_2015,Ottolina_2017,Moon_2019}.

\begin{figure*}
\includegraphics[width=15cm]{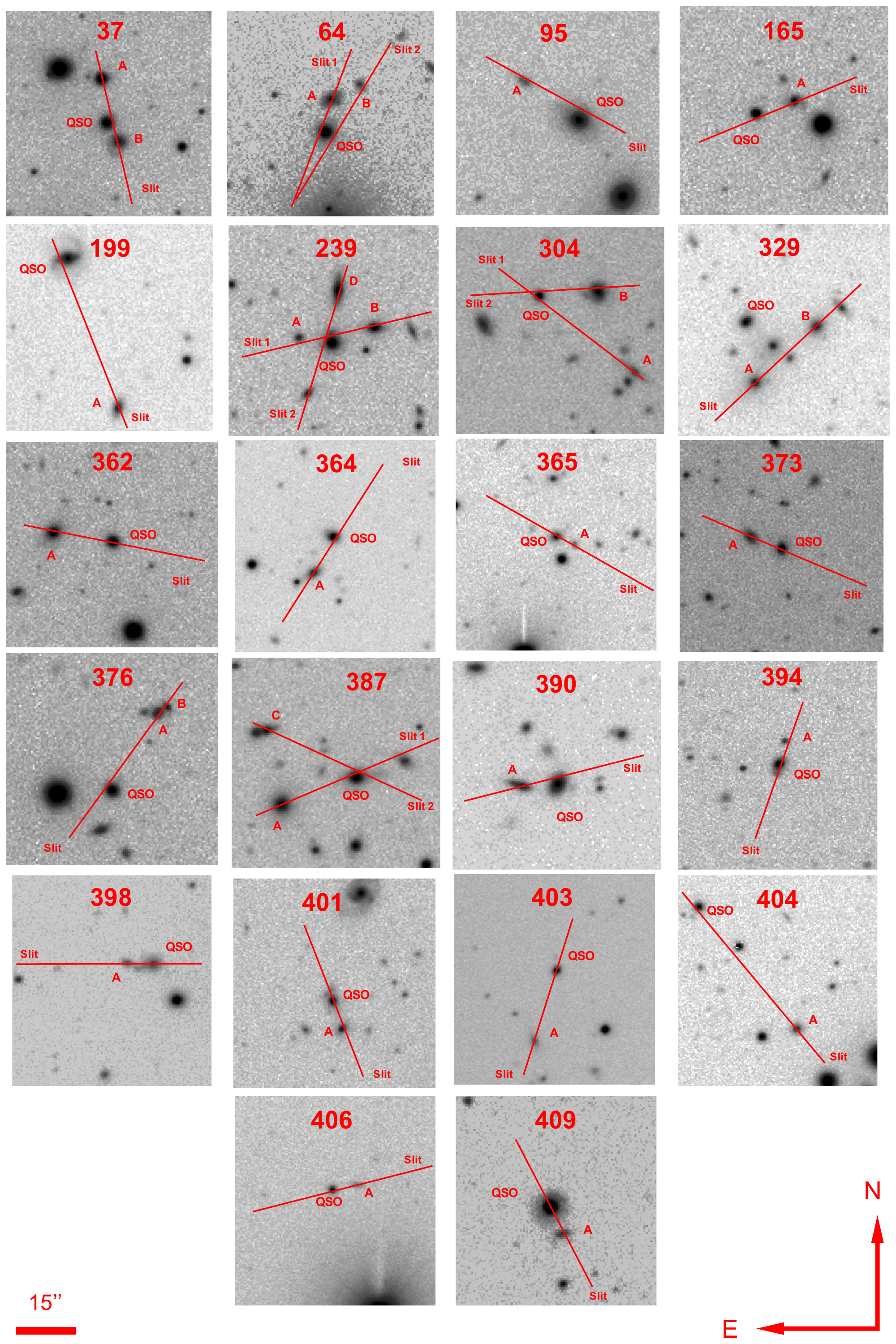}
\caption{SDSS $i$ band images of quasar fields. QSO numbers are from \protect\citetalias{Falomo_2014}. 
Companion galaxies are labelled by capital letters. The red lines indicate the slit positions.}
\label{fig:ima}
\end{figure*}

With these criteria we found 30 new QSO objects. 
Due to the observing conditions and telescope availability we could secure spectroscopy for 22 sources (i.e. $\sim$70\% of the whole sample). 
The main properties of the observed QSO are given in Table \ref{tab:sample} and in Fig. \ref{fig:ima} we show the fields of view of our targets.

\begin{table*}
 \centering
 \caption{The observed quasars and their companion galaxies.}
 \label{tab:sample}
 \begin{tabular}{rllrcrrrrrrrrr}
 \hline
 \multicolumn{1}{c}{Nr} &
 \multicolumn{1}{c}{Gal} &
 \multicolumn{1}{c}{SDSSJ} &
 \multicolumn{1}{c}{$z_{SDSS}$} &
 \multicolumn{1}{c}{Telescope} &
 \multicolumn{1}{c}{PD} &
 \multicolumn{1}{c}{PA} &
 \multicolumn{1}{c}{offset} &
 \multicolumn{1}{c}{u} &
 \multicolumn{1}{c}{g} &
 \multicolumn{1}{c}{r} &
 \multicolumn{1}{c}{i} &
 \multicolumn{1}{c}{z} &
 \multicolumn{1}{c}{FWHM} \\

& & & & & (kpc) & ($^{\circ}$) & (arcsec) & & & & & & (arcsec) \\
 \hline
37 & QSO & 211348.39$+$003722.1 & 0.3860 & NOT & & & 1.0 & 18.57 & 18.20 & 18.01 & 17.94 & 17.53 & 1.0\\
 & Gal. B & 211348.20$+$003717.6 & & NOT & 54 & 18 & & 22.60 & 20.69 & 19.45 & 18.85 & 18.44 & 1.0\\
64 & QSO & 215825.88$-$001804.7 & 0.3731 & NOT & & & 0.5 & 18.20 & 17.81 & 17.46 & 17.32 & 16.69 & 1.0\\
 & Gal. A & 215825.79$-$001756.8 & 0.4620 & NOT & 78 & 166 & & 22.93 & 21.66 & 19.82 & 18.99 & 18.59 & 1.0\\
 & Gal. B & 215825.34$-$001753.1 & & NOT & 137 & 150 & & 25.38 & 22.96 & 21.78 & 20.91 & 20.03 & 0.9\\
95 & QSO & 222909.81$+$002527.3 & 0.2276 & NOT & & & 1.6 & 18.52 & 18.36 & 17.82 & 17.34 & 17.25 & 1.3\\
 & Gal. A & 222910.69$+$002536.7 & & NOT & 89 & 60 & & 23.51 & 21.76 & 20.86 & 20.52 & 19.93 & 1.3\\
165 & QSO & 235926.25$-$004750.4 & 0.3961 & NOT & & & 1.0 & 20.01 & 19.56 & 19.46 & 19.13 & 18.61 & 1.2\\
 & Gal. A & 235925.64$-$004747.4 & & NOT & 100 & 102 & & 21.61 & 20.87 & 20.09 & 19.78 & 19.55 & 1.2\\
199 & QSO & 003723.49$+$000812.5 & 0.2517 & NOT & & & 1.0 & 18.23 & 18.15 & 17.81 & 17.59 & 17.32 & 1.2\\
 & Gal. A & 003722.69$+$000736.8 & & NOT & 231 & 20 & & 21.59 & 19.78 & 19.16 & 18.85 & 18.49 & 1.2\\
239 & QSO & 012050.94$-$001833.0 & 0.3492 & NOT & & & 2.2 & 19.20 & 18.87 & 18.48 & 18.36 & 17.85 & 1.0\\
 & Gal. A & 012051.45$-$001831.8 & & NOT & 69 & 97 & & 24.07 & 22.52 & 20.97 & 20.33 & 20.14 & 1.0\\
 & Gal. B & 012050.27$-$001829.4 & & NOT & 96 & 97 & & 24.24 & 21.37 & 19.68 & 19.06 & 18.63 & 1.0\\
 & Gal. D & 012050.84$-$001820.6 & & NOT & 112 & 164 & & 23.24 & 20.49 & 19.06 & 18.53 & 18.09 & 1.0\\
304 & QSO & 021046.47$-$004327.1 & 0.3864 & NOT & & & 0.9 & 19.45 & 19.26 & 19.37 & 19.28 & 18.70 & 1.5\\
 & Gal. A & 021044.96$-$004345.3 & & NOT & 294 & 49 & & 22.85 & 21.06 & 20.38 & 19.92 & 19.83 & 1.1\\
 & Gal. B & 021045.54$-$004326.1 & & NOT & 141 & 99 & & 21.24 & 20.03 & 19.03 & 18.63 & 18.21 & 1.5\\
 & Gal. C & 021047.96$-$004330.4 & & NOT & 228 & 99 & & 22.28 & 21.55 & 21.27 & 21.03 & 20.82 & 1.5\\
 & Gal. E & 021049.04$-$004250.5 & & NOT & 537 & 49 & & 22.34 & 21.61 & 20.35 & 19.87 & 19.41 & 1.1\\
329 & QSO & 024207.27$+$000038.7 & 0.3842 & NOT & & & 0.8 & 19.91 & 19.58 & 19.15 & 18.82 & 18.42 & 1.4\\
 & Gal. A & 024207.12$+$000024.6 & 0.3908 & NOT & 143 & 132 & & 23.11 & 21.04 & 19.23 & 18.56 & 18.17 & 1.0\\
 & Gal. B & 024206.17$+$000037.6 & & NOT & 166 & 132 & & 23.29 & 21.32 & 19.44 & 18.79 & 18.40 & 1.0\\
362 & QSO & 030745.95$+$000833.4 & 0.4270 & GTC & & & & 20.07 & 19.63 & 19.30 & 18.97 & 18.70 & 0.8\\
 & Gal.A & 030747.00$+$000835.9 & & GTC & 181 & 75 & & 23.36 & 20.88 & 19.30 & 18.69 & 18.20 & 0.8\\
364 & QSO & 030825.85$+$003054.3 & 0.3459 & GTC & & & & 19.26 & 18.93 & 18.47 & 18.31 & 17.80 & 1.4\\
 & Gal.A & 030826.21$+$003044.1 & & GTC & 102 & 60 & & 22.09 & 20.59 & 19.21 & 18.60 & 17.97 & 1.4\\
365 & QSO & 031039.34$-$004843.4 & 0.4569 & GTC & & & & 20.59 & 20.32 & 19.97 & 19.42 & 18.94 & 1.6\\
 & Gal.A & 031038.99$-$004845.7 & & GTC & 71 & 57 & & 23.95 & 22.63 & 21.44 & 21.18 & 21.18 & 1.6\\
 & Gal.B & 031033.54$-$004920.8 & & GTC & 1168 & 57 & & 23.68 & 21.54 & 20.98 & 20.60 & 20.48 & 1.6\\
373 & QSO & 032024.98$+$004418.0 & 0.4539 & GTC & & & & 20.20 & 19.69 & 19.41 & 19.05 & 18.71 & 1.7\\
 & Gal.A & 032025.58$+$004421.4 & & GTC & 118 & 67 & & 23.46 & 20.99 & 20.06 & 19.60 & 19.52 & 1.7\\
376 & QSO & 032234.07$+$002149.9 & 0.3483 & NOT & & & 1.2 & 19.41 & 19.25 & 18.96 & 18.87 & 18.40 & 0.9\\
 & Gal. A & 032233.35$+$002208.0 & 0.4109 & NOT & 189 & 144 & & 23.53 & 21.36 & 20.18 & 19.64 & 19.19 & 0.9\\
 & Gal. B & 032233.22$+$002209.4 & & NOT & 209 & 144 & & 24.09 & 24.18 & 21.84 & 21.51 & 20.70 & 0.9\\
387 & QSO & 032838.27$-$000341.6 & 0.3036 & NOT & & & 1.1 & 20.68 & 19.82 & 18.98 & 18.69 & 17.74 & 1.3\\
 & Gal. A & 032839.46$-$000348.1 & 0.3013 & NOT & 145 & 113 & & 21.98 & 20.45 & 18.77 & 18.25 & 17.79 & 1.1\\
 & Gal. C & 032837.52$-$000337.8 & & NOT & 183 & 60 & & 23.12 & 22.24 & 19.97 & 19.46 & 18.95 & 1.3\\
 & Gal. Y & 032835.24$-$000409.2 & & NOT & 405 & 60 & & 20.71 & 19.09 & 18.19 & 17.70 & 17.31 & 1.3\\
390 & QSO & 033156.88$+$002605.2 & 0.2369 & NOT & & & 1.3 & 20.06 & 18.95 & 17.94 & 17.44 & 17.18 & 1.2\\
 & Gal. A & 033157.44$+$002604.7 & & NOT & 48 & 103 & & 22.05 & 20.55 & 19.38 & 18.81 & 18.36 & 1.2\\
394 & QSO & 033305.96$+$005735.9 & 0.3118 & GTC & & & & 19.57 & 19.32 & 18.95 & 18.86 & 18.27 & 2.0\\
 & Gal.A & 033305.84$+$005743.1 & & GTC & 58 & 163 & & 23.32 & 23.06 & 21.76 & 20.95 & 20.99 & 2.0\\
 & Gal.C & 033306.97$+$005649.4 & & GTC & 385 & 163 & & 22.13 & 21.80 & 21.22 & 20.91 & 20.29 & 2.0\\
398 & QSO & 033431.31$+$005121.5 & 0.4295 & GTC & & & & 20.35 & 19.90 & 19.22 & 18.74 & 18.46 & 2.4\\
 & Gal.A & 033431.61$+$005120.0 & & GTC & 84 & 81 & & 23.71 & 23.78 & 22.16 & 21.60 & 21.45 & 2.4\\
401 & QSO & 033627.43$+$000442.9 & 0.3957 & GTC & & & & 20.49 & 20.04 & 19.52 & 19.11 & 18.70 & 3.6\\
 & Gal.A & 033627.24$+$000434.8 & & GTC & 89 & 22 & & 22.77 & 22.76 & 20.72 & 19.43 & 18.81 & 3.6\\
 & Gal.B & 033625.24$+$000318.1 & & GTC & 946 & 22 & & 21.50 & 20.30 & 19.73 & 19.51 & 19.47 & 3.6\\
403 & QSO & 033718.81$+$003303.7 & 0.4371 & GTC & & & & 20.93 & 20.18 & 19.27 & 18.82 & 18.52 & 1.5\\
 & Gal.A & 033719.24$+$003242.8 & & GTC & 256 & 162 & & 23.16 & 21.58 & 19.96 & 19.35 & 19.06 & 1.5\\
404 & QSO & 033852.90$+$001904.7 & 0.4594 & GTC & & & & 20.07 & 19.76 & 19.49 & 19.21 & 18.89 & 1.7\\
 & Gal.A & 033851.00$+$001828.3 & & GTC & 574 & 118 & & 23.91 & 21.89 & 20.23 & 19.51 & 19.04 & 1.7\\
406 & QSO & 033938.59$-$003625.6 & 0.4911 & GTC & & & & 19.86 & 19.58 & 19.53 & 19.29 & 19.11 & NA\\
 & Gal.A & 033938.11$-$003624.1 & & GTC & 99 & 100 & & 23.23 & 22.85 & 21.22 & 20.84 & 20.14 & NA\\
409 & QSO & 034226.50$-$000427.1 & 0.3765 & GTC & & & & 17.43 & 17.36 & 17.40 & 17.46 & 17.00 & 2.0\\
 & Gal.A & 034226.30$-$000433.4 & & GTC & 68 & 26 & & 22.93 & 21.71 & 20.27 & 19.82 & 19.23 & 2.0\\
 \hline
 \end{tabular}
 \begin{list}{}{}
 \item[]Column (1) \citetalias{Falomo_2014} catalog identification number. 
 Column (2) Object label. 
 Columns (3) and (4) SDSS DR7 name and redshift. 
 Column (5) Telescope used. 
 Column (6) Observed PD of the companion from QSO. 
 Column (7) Position angle (PA) with respect to the QSO. 
 Column (8) Slit offset from the QSO centre. 
 Columns (9) to (13) SDSS DR7 $u$, $g$, $r$, $i$ and $z$ magnitudes. 
 Column (14) Seeing, full width half maximum (FWHM) during the observations.
 \end{list} 
\end{table*}

\section{Observations and Data Analysis}

We observed 22 new QSO fields for which we secured one or more long slit spectra of the companions and of the quasar.
Data for 11 QSO were obtained at the 10.4 m GTC at the Roque de Los Muchachos with the spectrograph OSIRIS (Optical System for Imaging and low-Intermediate-Resolution Integrated Spectroscopy; \citealt{Cepa_2003}) covering the spectral range 4100-9000 {\AA}. 
We adopted the grism R2500 and a slit width of 1.2 arcsec, yielding an effective spectral resolution of $R=$ 800.
The GTC observations were carried out in December 2017 and in January 2018. 

For the other 11 QSO the data were obtained at the NOT telescope equipped with the Alhambra Faint Object Spectrograph and Camera (ALFOSC). 
In this case we used the grism \#7 and a slit width of 1.3 arcsec, covering the wavelength range of 3650-7110 {\AA} and yielding a spectral resolution $R=$ 500. 
These observations were collected in October 2017 and September 2018. 

On average an exposure time of one hour was adopted for each observation.
We positioned the slit to intercept both the QSO and the candidate companion galaxy. 
In some cases the slit serendipitously crossed other objects, all these spectra have been reduced and added to our results. 
For this reason objects at larger than 250 kpc PDs are listed in Table \ref{tab:res}.
For all NOT targets we were able to place the slit slightly offset from the nucleus of the QSO in order to optimise the signal from the host galaxy. 
The list of the observed targets is given in Table \ref{tab:sample}.

All of the reduction steps were performed using \textsc{iraf}\footnote{
\textsc{iraf} is distributed by the National Optical Astronomy Observatory, which is operated by the Association of Universities for Research in Astronomy (AURA) under cooperative agreement with the National Science Foundation
} 
procedures.
Multiple exposures were combined with cosmic ray rejection option.
The instrumental signatures were removed by applying bias subtraction and flat field correction. 
The wavelength calibration was performed using HeNe arc frames with accuracy of 0.2 {\AA}. 
After background subtraction, 1D spectra were extracted.

The relative flux calibration was performed using a sensitivity function derived from a standard star observed during the same run.
Since using a narrow slit might result in a loss of a fraction of the light of the companion galaxies and of the QSO, we performed absolute flux calibration. 
For the companion galaxies, we normalised the continuum to the flux matching the SDSS $r$-band magnitudes (DR7).
If detected, the host galaxy flux was normalised to SDSS $i$-band magnitudes as derived in our previous work (\citetalias{Falomo_2014}, \citealt{Bettoni_2015}).
The calibrated final spectra of the targets are given in Appendix \ref{sec:appendix_spectra}.

The redshifts of the observed targets were calculated using the {\it emsao} and {\it xcsao} routines of the \textsc{rvsao} package \citep{Kurtz_1998}.
For the {\it xcsao} task (i.e. pure absorption line spectra), we used a synthetic reference stellar spectrum of a K\,\textsc{iii} star \citep{Jacoby_1984}.

\begin{table*}
\centering
 \caption{[O\,\textsc{ii}] emission-line measurements.}
 \label{tab:res}
\makebox[\textwidth][c]{
 \begin{tabular}{cllccccc}
 \hline
 \multicolumn{1}{c}{Nr} &
 \multicolumn{1}{c}{Obj} &
 \multicolumn{1}{c}{z$_{our}$} &
 \multicolumn{1}{c}{log(L$_{O\,\textsc{ii}}$)} &
 \multicolumn{1}{r}{log(L$_{O\,\textsc{ii}}$/L$_{\odot}$)} &
 \multicolumn{1}{c}{SFR} &
 \multicolumn{1}{c}{sSFR} &
 \multicolumn{1}{c}{log(M$*$)} \\
& & & (erg s$^{-1}$) & & (M$_{\odot}$ yr$^{-1}$) & (yr$^{-1} \times 10^{-10}$) & (M/M$_{\odot}$) \\
 \hline
37 & QSO & 0.3864 & 41.95 $\pm$ 0.02 & 8.37 & -- & -- & -- \\
 & \textbf{Gal. B} & \textbf{0.3862} & 41.80 $\pm$ 0.04 & 8.22 & 55.2 $\pm$ 4.6 & 3.5 & 11.19 \\

64 & QSO & 0.3737 & 42.12 $\pm$ 0.01 & 8.54 & -- & -- & -- \\
 & Gal. A & 0.4612 & 41.30 $\pm$ 0.08 & 7.71 & 18.7 $\pm$ 3.4 & 0.3 & 11.78 \\
 & Gal. B & 0.5016 & 40.83 $\pm$ 0.06 & 7.24 & 5.3 $\pm$ 0.7 & 0.6 & 10.93 \\

95 & QSO & 0.2272 & 41.28 $\pm$ 0.07 & 7.70 & 14.7 $\pm$ 2.2 & 1.8 & 10.91 \\
 & Gal. A & 0.4729 ? & 41.34 $\pm$ 0.03 & 7.76 & 9.3 $\pm$ 0.7 & 3.8 & 10.39 \\

165 & QSO & 0.3965 & 41.28 $\pm$ 0.02 & 7.70 & -- & -- & -- \\
 & \textbf{Gal. A} & \textbf{0.3957} & 41.49 $\pm$ 0.02 & 7.90 & 19.9 $\pm$ 1.0 & 4.0 & 10.69 \\

199 & QSO & 0.2510 & 41.00 $\pm$ 0.05 & 7.41 & 4.1 $\pm$ 0.4 & 1.7 & 10.39 \\
 & Gal. A & 0.0690 & 39.90 $\pm$ 0.02 & 6.32 & 0.1 $\pm$ 0.0 & 0.5 & 9.14 \\

239 & QSO & 0.3495 & 41.83 $\pm$ 0.07 & 8.24 & -- & -- & -- \\
 & \textbf{Gal. A} & \textbf{0.3495} & 40.68 $\pm$ 0.05 & 7.10 & 1.9 $\pm$ 0.2 & 0.8 & 10.36 \\
 & Gal. B & 0.3454 & -- & -- & -- & -- & 11.07 \\
 & Gal. D & 0.2653 & -- & -- & -- & -- & 11.07 \\

304 & QSO & 0.3857 & 41.30 $\pm$ 0.03 & 7.72 & -- & -- & -- \\
 & Gal. A & 0.2837 & 41.37 $\pm$ 0.01 & 7.78 & 4.2 $\pm$ 0.1 & 5.0 & 9.93 \\
 & Gal. B & 0.2821 & 41.05 $\pm$ 0.05 & 7.47 & 7.9 $\pm$ 0.9 & 1.3 & 10.77 \\
 & Gal. C & 0.1697 & 40.49 $\pm$ 0.05 & 6.91 & 0.3 $\pm$ 0.0 & 4.0 & 8.83 \\
 & Gal. E & 0.3810 & 40.89 $\pm$ 0.07 & 7.31 & 5.1 $\pm$ 0.8 & 1.0 & 10.70 \\

329 & QSO & 0.3844 & 41.54 $\pm$ 0.05 & 7.96 & -- & -- & -- \\
 & Gal. A & 0.3912 & 41.47 $\pm$ 0.06 & 7.89 & 27.8 $\pm$ 3.8 & 0.6 & 11.66 \\
 & Gal. B & 0.3947 & -- & -- & -- & -- & 11.59 \\

362 & QSO & 0.4277 & 40.55 $\pm$ 0.05 & 6.97 & -- & -- & -- \\
 & Gal. A & 0.2777 & 41.72 $\pm$ 0.03 & 8.14 & 43.2 $\pm$ 3.0 & 4.2 & 11.01 \\

364 & QSO & 0.3461 & 41.38 $\pm$ 0.01 & 7.80 & -- & -- & -- \\
 & \textbf{Gal. A} & \textbf{0.3467} & 42.39 $\pm$ 0.01 & 8.80 & 215.5 $\pm$ 5.0 & 13.1 & 11.22 \\

365 & QSO & 0.4578 & 41.31 $\pm$ 0.02 & 7.73 & -- & -- & -- \\
 & \textbf{Gal. A} & \textbf{0.4567} & 41.54 $\pm$ 0.01 & 7.96 & 11.4 $\pm$ 0.2 & 6.5 & 10.25 \\
 & Gal. B & 0.1078 & -- & -- & -- & -- & 8.60 \\

373 & QSO & 0.4543 & 40.99 $\pm$ 0.02 & 7.41 & -- & -- & -- \\
 & Gal. A & 0.1868 & -- & -- & -- & -- & 9.90 \\

376 & QSO & 0.3491 & 40.91 $\pm$ 0.10 & 7.33 & -- & -- & -- \\
 & Gal. A & 0.4106 & 41.20 $\pm$ 0.05 & 7.62 & 12.9 $\pm$ 1.6 & 1.3 & 10.99 \\
 & Gal. B & 0.4105 & 40.31 $\pm$ 0.09 & 6.73 & 1.5 $\pm$ 0.3 & 0.2 & 10.81 \\

387 & QSO & 0.3034 & 41.87 $\pm$ 0.02 & 8.29 & -- & -- & -- \\
 & Gal. A & 0.3015 & 41.12 $\pm$ 0.06 & 7.54 & 11.9 $\pm$ 1.6 & 0.6 & 11.32 \\
 & Gal. C & 0.3004 & -- & -- & -- & -- & 11.06 \\
 & Gal. Y & 0.0919 & 40.17 $\pm$ 0.05 & 6.59 & 0.4 $\pm$ 0.0 & 0.3 & 10.12 \\

390 & QSO & 0.2369 & -- & -- & -- & -- & 10.73 \\
 & \textbf{Gal. A} & \textbf{0.2370} & 40.27 $\pm$ 0.08 & 6.68 & 1.4 $\pm$ 0.2 & 0.2 & 10.84 \\

394 & QSO & 0.3123 & 40.84 $\pm$ 0.02 & 7.26 & -- & -- & -- \\
 & \textbf{Gal. A} & \textbf{0.3135} & 40.69 $\pm$ 0.03 & 7.10 & 0.6 $\pm$ 0.0 & 1.3 & 9.68 \\
 & Gal. C & 0.0888 & -- & -- & -- & -- & 8.52 \\

398 & QSO & 0.4297 & 41.51 $\pm$ 0.03 & 7.93 & -- & -- & -- \\
 & Gal. A & 0.4381 ? & 40.91 $\pm$ 0.08 & 7.33 & 7.2 $\pm$ 1.3 & 0.5 & 11.17 \\

401 & QSO & 0.3960 & 41.18 $\pm$ 0.02 & 7.60 & -- & -- & -- \\
 & Gal. A & 0.3750 ? & 40.50 $\pm$ 0.10 & 6.92 & 2.8 $\pm$ 0.6 & 0.2 & 11.23 \\
 & Gal. B & 0.3389 ? & 40.55 $\pm$ 0.10 & 6.97 & 1.8 $\pm$ 0.4 & 0.6 & 10.48 \\

403 & QSO & 0.4370 & 41.67 $\pm$ 0.01 & 8.09 & -- & -- & -- \\
 & Gal. A & 0.3076 & -- & -- & -- & -- & 10.71 \\

404 & QSO & 0.4601 & 41.50 $\pm$ 0.02 & 7.92 & -- & -- & -- \\
 & Gal. A & 0.4263 & 42.18 $\pm$ 0.04 & 8.60 & 138.5 $\pm$ 11.9 & 6.0 & 11.36 \\

406 & QSO & 0.4918 & 40.83 $\pm$ 0.02 & 7.25 & -- & -- & -- \\
 & Gal. A & 0.2486 & 41.14 $\pm$ 0.05 & 7.56 & 3.8 $\pm$ 0.4 & 2.6 & 10.16 \\

409 & QSO & 0.3766 & 40.97 $\pm$ 0.02 & 7.39 & -- & -- & -- \\
 & Gal. A & 0.4678 ? & -- & -- & -- & -- & 11.32 \\
 \hline
 \end{tabular}}
 \newline
 \raggedright{Column (1) QSO identification number from \protect\citetalias{Falomo_2014} catalog. 
 Column (2) Redshift measured in this study (true companion galaxies bolded). 
 Columns (3) and (4) [O\,\textsc{ii}] emission line luminosities (error is based on the flux measurement error in \textsc{iraf} $splot$ task; \citealt{Wright_2006}). 
 Columns (5) and (6) SFR and sSFR ($sSFR = SFR / M_{\star}$). 
 Column (7) Stellar mass \citep{Gilbank_2010,Bettoni_2015}.}
\end{table*}

\section{Results}
In this work we use the following definitions, as in Paper \citetalias{Bettoni_2017}.
Close companion galaxies are galaxies which could potentially merge with the quasar and in this work we consider them to be located within a few hundred kpc (PD) from the quasar.
We consider them candidate companion galaxies until the redshift is determined (note that we use the terms "candidate companion galaxies", "companion galaxies" and "companions" interchangeably).
Physically associated companion galaxies (or in short 'associated galaxies') are those where the radial velocity difference with the quasar is less than 400 km s$^{-1}$.
Above that limit, the candidate companion galaxies are considered to be physically non-associated to the quasar.

The new spectroscopic observations provide data for 22 QSO and 35 candidate close companions for which we could measure their redshift (see Table \ref{tab:res}).
The individual notes for each field are given in the Appendix \ref{sec:fields} and the final spectra are given in the Appendix \ref{sec:appendix_spectra}.
From this new dataset we found that only in 7 fields out of 22 ($\sim$32\%) the close companion object is associated with the QSO.
Moreover, at odds to what is found in Paper \citetalias{Bettoni_2017} no cases of multiple companions are found (note that few QSO (e.g. 239) had more candidate companions which however were not included in this study due to observing constraints).
Combining this study with that reported in Paper \citetalias{Bettoni_2017}, the total sample is composed of 34 QSO at redshift of 0.2\textless z\textless0.5 (see Table \ref{tab:statistics}).
The total number of candidate close companions is thus 56 (Fig. \ref{fig:histograms}).
The mean redshift of the whole sample of QSO is $0.31\pm0.10$.
On average the uncertainty in measured difference of radial velocity is $83\pm56$ km s$^{-1}$.

Note that there are a number of differences between the two datasets that need to be considered when comparing the results (see Table \ref{tab:statistics}). 
The objects in Paper \citetalias{Bettoni_2017} have on average a lower redshift (mean z $\sim$ $0.20\pm$0.05) with respect to the sources in this work (mean z $\sim$ $0.37\pm0.07$).
In this work, the candidate companion galaxies have a median SDSS $r$-band apparent magnitude of $20.09\pm0.98$, which is about half a magnitude fainter than the median apparent magnitude of Paper \citetalias{Bettoni_2017} ($19.58\pm1.31$).
On the other hand, the average absolute magnitude of the companion galaxies in this work is about half a magnitude brighter than the companions in Paper \citetalias{Bettoni_2017} (see Table \ref{tab:statistics}).
Paper \citetalias{Bettoni_2017} sample has a small number of faint outliers.
Otherwise, even though the redshift sampled is higher in this work as compared to Paper \citetalias{Bettoni_2017}, the distribution of the absolute magnitudes is quite similar.
The PDs of candidate companions in this work are concentrated within the 200 kpc radius from the QSO, with only a few outliers between 200 kpc and $\sim$1 Mpc (see Section 2 for explanation of companions with higher PD).
Probably the most important difference is that the median PD of companions in the new dataset is more than two times larger than that of companions reported in Paper \citetalias{Bettoni_2017} (see Table \ref{tab:statistics}).

\begin{table*}%
 \centering
 \caption{Statistical overview of the results.}
 \label{tab:statistics}
 \begin{tabular}{lcccccccccc}
 \hline
 \multicolumn{1}{c}{Sample} &
 \multicolumn{1}{c}{$N_{QSO}$} &
 \multicolumn{1}{c}{$N_{C}$} &
 \multicolumn{1}{c}{$\overline{z}$} &
 \multicolumn{1}{c}{$\widetilde{PD}$} &
 \multicolumn{1}{c}{$\overline{M_{r}}$} &
 \multicolumn{1}{c}{$\overline{m_{r}}$} &
 \multicolumn{1}{c}{$N_{O\,\textsc{ii}}$} &
 \multicolumn{1}{c}{$\overline{L_{O\,\textsc{ii}}}$} &
 \multicolumn{1}{c}{$\widetilde{SFR}$} &
 \multicolumn{1}{c}{$\overline{sSFR}$} \\
 & & & & (kpc) & & & & (erg s$^{-1}$) & (M$_{\odot}$ yr$^{-1}$) & (yr$^{-1} \times 10^{-10}$) \\
 \hline
Paper \citetalias{Bettoni_2017} & 12 & 8 & 0.20$\pm$0.05 & 62.7$^{+19.7}_{-27.2}$ & -20.21$\pm$1.60 & 19.39$\pm$1.44 & 5 & 40.50$\pm$0.78 & 1.0$^{+5.8}_{-0.5}$ & 4.0$\pm$6.7 \\[0.2cm]
This work & 22 & 7 & 0.37$\pm$0.07 & 141.0$^{+88.5}_{-52.0}$ & -20.71$\pm$1.38 & 20.33$\pm$0.98 & 7 & 41.26$\pm$0.69 & 11.4$^{+26.1}_{-9.7}$ & 4.2$\pm$4.1 \\[0.2cm]
Whole & 34 & 15 & 0.31$\pm$0.10 & 97.5$^{+84.0}_{-31.2}$ & -20.52$\pm$1.48 & 19.76$\pm$1.36 & 12 & 40.88$\pm$0.83 & 4.3$^{+10.1}_{-3.4}$ & 4.1$\pm$5.5 \\
 \hline
 \end{tabular}
 \newline
 \raggedright{Column (1) Run. 
 Column (2) Number of QSO fields observed. 
 Column (3) Number of QSO with at least one associated companion. In this work, from NOT targets 4/11 QSO had an associated companion and for GTC targets 3/11.
 Column (4) Mean redshift of QSO. 
 Column (5) Median PD of companions. 
 Column (6) Mean SDSS M$_r$ absolute magnitude of companions at their redshifts. 
 Column (7) Mean SDSS m$_r$ of associated companion galaxies. 
 Column (8) Number of associated companions with [O\,\textsc{ii}] emission line. 
 Columns (9-11) Mean [O\,\textsc{ii}] luminosity, median SFR, and mean sSFR of associated companion galaxies with [O\,\textsc{ii}] emission line.
 The mean values are reported with standard deviations and the median with upper and lower quartile errors.}
\end{table*}

\subsection{Close companion galaxies}

If the galaxy and quasar have a difference in radial velocities $\Delta$V\textless400 km s$^{-1}$, we consider them to be physically associated, as in Paper \citetalias{Bettoni_2017}.
This criterion for the line of sight velocity establishes a reasonable balance to include a sufficient number of companions which are likely to be within the gravitational interaction radius of the quasar.
Many previous studies used a similar cutoff based on distribution statistics of targets and companions \citep{Patton_2000,Lambas_2003,Ellison_2008, Keenan_2014, Nazaryan_2014b, Robotham_2014, Moon_2019} and the physics background on the identification of companions is discussed in \citealt{Valtonen_1986,Byrd_1987}.
This criterion is satisfied for 7 (31.8\%) candidate companions in this work (see Table \ref{tab:res}).
The other companions turned out to be either foreground or background galaxies.
The slits also intercepted five objects which turned out to be Galactic stars and are excluded from the results.

\begin{figure}
	\subfigure{\includegraphics[width = \columnwidth]{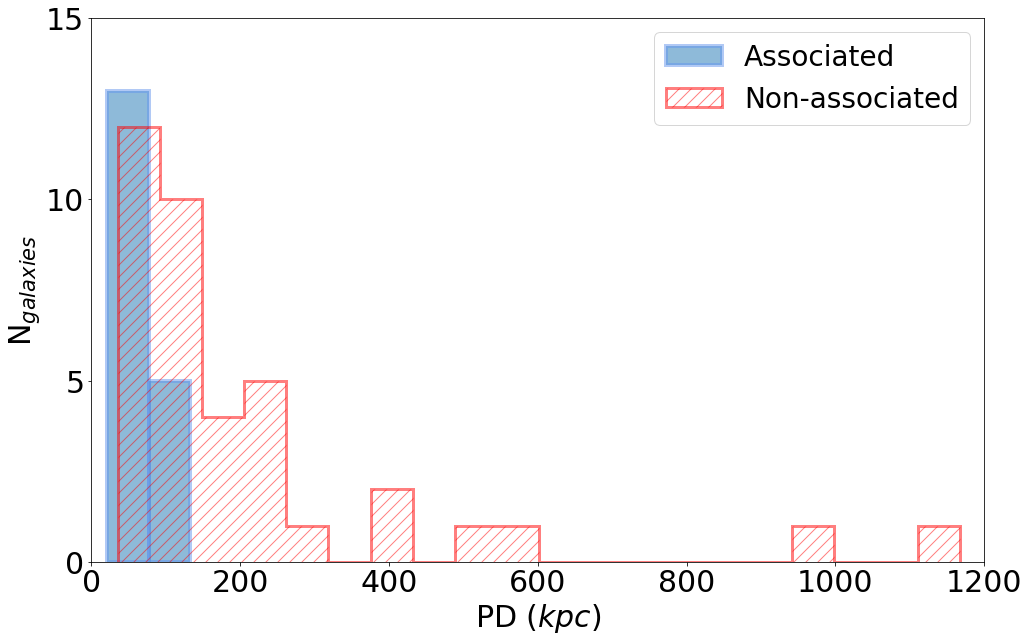}}
	\subfigure{\includegraphics[width = \columnwidth]{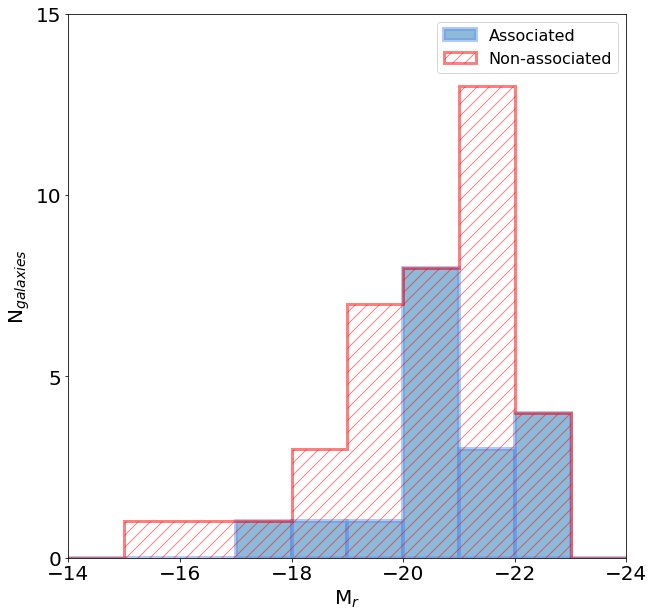}}
\caption{Distribution of PD ({\it top}) and of absolute magnitudes ({\it bottom}) for the associated ({\it blue bars}) and non-associated companion galaxies ({\it red hashed histogram}) in the whole sample.}
\label{fig:histograms}
\end{figure}

Together with data from Paper \citetalias{Bettoni_2017}, the whole sample we have is composed by 34 QSO; 15 objects (44\%) had at least one associated companion, while the remaining 19 QSO (56\%) did not have any associated companions.
The median redshift of all QSO with at least one companion is 0.25 (for Paper \citetalias{Bettoni_2017} 0.21 and for this work 0.35).
In Paper \citetalias{Bettoni_2017}, 11 (52\%) companions are associated to the QSO, while in this work only 7 (20\%).
In total, we had 56 spectra of galaxies, out of which 18 (32.1\%) turned out to be associated to the QSO (Fig. \ref{fig:histograms}).
Note that from Paper \citetalias{Bettoni_2017}, two QSO had more than a single associated galaxy.

\begin{figure}
 \includegraphics[width = \columnwidth]{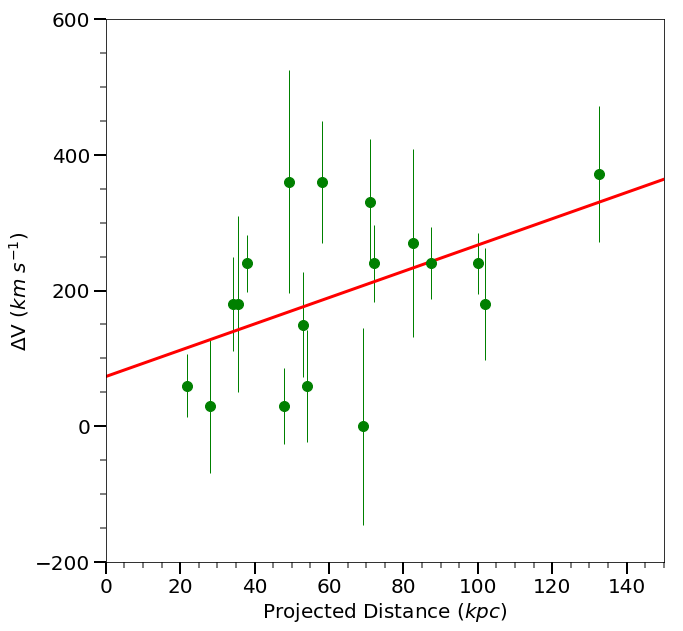}
\caption{Radial velocity difference between the QSO and the associated companion galaxy as a function of the PD of the companion.
The solid red line is the formal best-fit; $\Delta V = 1.935 \times PD + 73.407$, correlation coefficient $r=0.469$.}
\label{fig:radial_velocity_plot}
\end{figure}

Considering only the population of associated companions, the average SDSS $r$ apparent magnitude of associated companions in this work is about one magnitude fainter than in Paper \citetalias{Bettoni_2017}.
In this work, the SDSS $r$ apparent magnitude of associated companion galaxies does not deviate much from the mean, while in Paper \citetalias{Bettoni_2017} several galaxies extend to the faint end of the distribution.
However, in the whole sample, the mean absolute magnitude of associated companions ($-20.71\pm1.39$) and non-associated companions ($-20.43\pm1.51$) is similar (Fig. \ref{fig:histograms}).
Furthermore, all confirmed companions are found to be within 150 kpc PD (Fig. \ref{fig:radial_velocity_plot}) and are generally projected closer to the quasar and have smaller scatter than non-associated galaxies which are just randomly projected along the line of sight (Fig. \ref{fig:histograms}).
Associated companions with shorter PDs from QSO have smaller radial velocity differences (Fig. \ref{fig:radial_velocity_plot}).

Although we present new data for about twice as many QSO compared to Paper \citetalias{Bettoni_2017}, we found a similar number of QSO with a physically associated companion.
However, the difference is not statistically significant.
To rule out a selection bias, thus we looked at PD, apparent and absolute magnitude distributions of both subsamples of candidate companion galaxies.
Nevertheless, no obvious differences have been distinguished.

We observed that a sizable fraction of quasars (44\% of the whole sample) had an associated companion.
For the rest of the quasars in our sample we did not find an associated companion galaxy.
Even though for some quasars we did not obtain a redshift for each possible companion (see Fig. \ref{fig:ima} and \citealt{Bettoni_2017}{}) due to the observing conditions, we obtained redshifts for 56 candidate companions, out of which 18 were associated to the quasar.
Within the merger scenario, the associated companions are merger candidates; thus one explanation for quasars lacking merger candidates is that the merger is in an advanced stage, i.e. the quasar and the companion galaxy have already merged \citep{Byrd_2001,Hopkins_2005_buried}.
Another possibility is that for these quasars without an associated companion the triggering of the SMBH fueling was driven by secular evolution processes such as slower gas accretion, instabilities in spiral arms or bars.

\subsection{Star formation from [O \textsc{ii}] emission lines}
Since either a merger or gas accretion can trigger star formation, we obtained the SFR in our sample of quasars and close companions.
We measured the [O\,\textsc{ii}] 3727 {\AA} forbidden line to derive an estimate of SFR, as it is a strong emission line in the optical spectrum at our redshifts and since the [O\,\textsc{ii}] and H \textbeta{} emission lines are well correlated \citep{Gallagher_1989,Kennicutt_1998,Kennicutt_2012}.
We estimated the flux of the [O\,\textsc{ii}] 3727 {\AA} line by fitting a single Gaussian profile.
The [O\,\textsc{ii}] emission line measurements are in Table \ref{tab:res}, showing the calculated luminosities, star formation rates and specific star formation rates (sSFR).

To calculate the SFR, we employed the equation by \cite{Gilbank_2010} (equation 8; also see \citealt{Gilbank_2011}), which includes a correction for the dust extinction and metallicity via a mass-term, as described in Paper \citetalias{Bettoni_2017},

\begin{equation}
SFR_{emp,corr} = \frac{SFR_0}{a \ tanh [(x-b)/c] + d}.
\end{equation}

Here, $x= log_{10}( M_{\star} / M_{\astrosun})$, $a = -1.424$, $b=9.827$, $c=0.572$, $d=1.700$.
$SFR_0 / (M_{\astrosun} \cdot yr^{-1}) = L([OII]) / 3.80 \cdot 10^{40} erg \cdot s^{-1}$.
The $SFR_0$ quantity is the nominal [O\,\textsc{ii}] SFR, which is based on equation 7 from \cite{Gilbank_2010}, and where $r_{lines} = 0.5$ and $H_{\alpha}$ extinction term is assumed to be 1 mag (and also see \citealt{Kennicutt_1998}).
The stellar mass for the quasar hosts is taken from \citealt{Bettoni_2015} and that of companions is calculated following the equation (1) in \cite{Gilbank_2010}, $log_{10}(M_{\star}/M_{\astrosun}) = 0.480(g-r)_0 - 0.475M_z - 0.08$.

In this work we detected [O\,\textsc{ii}] 3727 {\AA} emission line in 47 targets (QSO and candidate companions) out of the total number of 57 (82.4\%).
This emission line is present in the majority (95.5\%) of QSO host spectra and in 26 candidate companions (74.3\%).
It is worth to note that in all cases of the new associated companion galaxies we found an [O\,\textsc{ii}] emission line in their spectra.
In Paper \citetalias{Bettoni_2017} in the sample of 21 galaxies (both associated and not associated galaxies), we detect emission only in 11 (53\%). 

Considering the whole sample of associated companion galaxies, the [O\,\textsc{ii}] emission line is present in 12 (67\%).
For the associated companion galaxies in this work, the Equivalent Width (EW) range (4-63 \AA) is similar to Paper \citetalias{Bettoni_2017}.
For consistency, we also measured the [O\,\textsc{ii}] emission in non-associated companions in this work; it is present in a large fraction (66\%) of them.

In this work, we found that the [O\,\textsc{ii}] luminosity ranges from $10^{40}$ to $2 \times 10^{42}$ erg s$^{-1}$. The SFR for candidate companion galaxies ranges from $\sim$0.1 to $\sim$55 M$_{\odot}$ yr$^{-1}$, except for Gal 364A and Gal 404A whose SFR is upwards of $10^2$ M$_{\odot}$ yr$^{-1}$. 
The SFR for associated companions ranges from $\sim$0.6 to $\sim$215 M$_{\odot}$ yr$^{-1}$.
Galaxy A associated to QSO 364 exhibits the highest SFR of all observed companion galaxies .

We were able to find quasar host masses for three QSO in this work from \citealt{Bettoni_2015}, but only two had SFR based on [O\,\textsc{ii}] emission (Table \ref{tab:res}). 
The sSFR for five associated companions (out of seven) is higher than 1 Gyr$^{-1}$, and could be due to a recent star formation activity.
The median SFR of associated galaxies in this work is higher than that found for the subsample in Paper \citetalias{Bettoni_2017}.\ 
However, we note that there are two associated companions in the new subsample that exhibit a very high SFR.
Without these two companions the distribution of SFR for the associated galaxies in the two subsamples is very similar.
Moreover, the sSFR, which is a better indicator of how the star formation of a galaxy is contributing to the build up of the stellar mass ( e.g. \citealt{Feulner_2005}), is comparable between the two subsamples.

The mean [O\,\textsc{ii}] luminosity of associated ($40.88\pm0.83$ erg s$^{-1}$) and non-associated ($40.83\pm0.66$ erg s$^{-1}$) galaxies from the whole sample are very similar.
The median SFR of associated galaxies ($4.3^{+10.1}_{-3.4}$ M$_{\odot}$ yr$^{-1}$) does not differ much from that of non-associated galaxies ($5.1^{+6.8}_{-3.5}$ M$_{\odot}$ yr$^{-1}$).
The SFR of associated galaxies are in agreement with reported SFR values for emission-line galaxies at comparable redshift ranges \citep{Mouhcine_2006}.

\begin{figure}
 \includegraphics[width = \columnwidth]{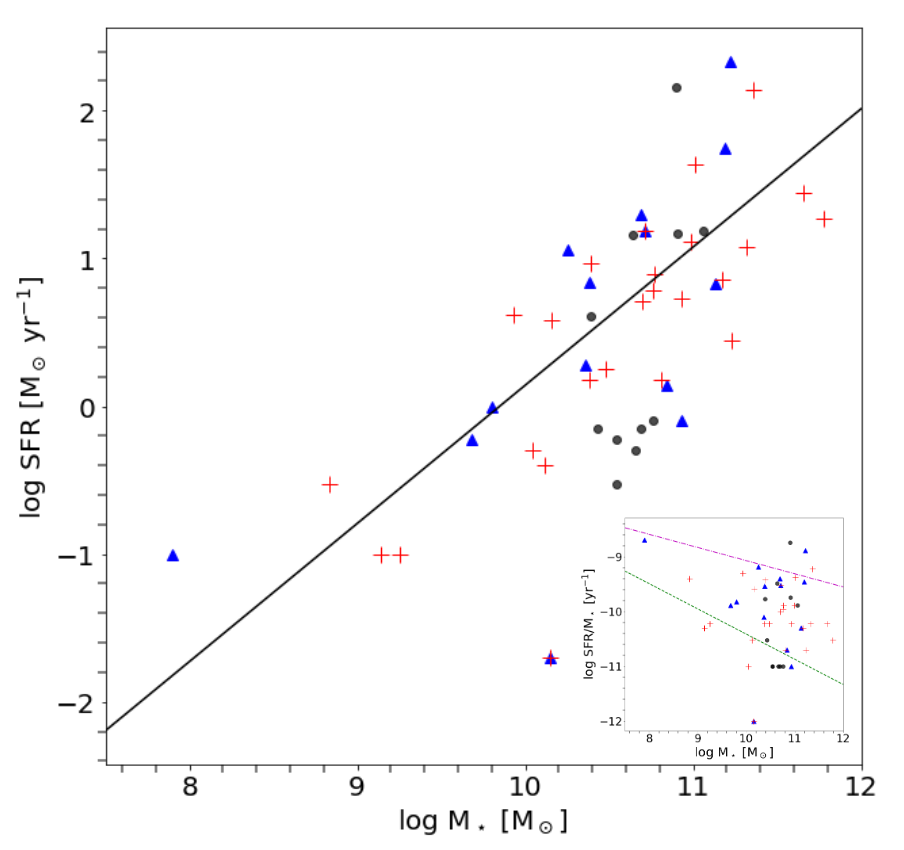}
\caption{SFR versus galaxy mass for the whole sample of associated (blue triangles) and non-associated galaxies (red crosses). 
Only quasar host galaxies (black circles) with available masses \citep{Bettoni_2015} are shown. 
For reference we plot the main sequence of star formation for the SDSS star-forming galaxies from \citealt{Duarte_2017}. 
In the inset we show the sSFR versus the stellar mass of the galaxies. 
We overplot z$\sim$0 (green dashed line) and z$\sim$1 (magenta dash dot line) relations from \citealt{Osborne_2020}.
}
\label{fig:main_sequence}
\end{figure}

On the other hand the mean of the sSFR of associated galaxies ($4.1\pm5.5$ yr$^{-1} \times 10^{-10}$) is higher than that of non-associated galaxies ($1.5\pm1.7$ yr$^{-1} \times 10^{-10}$).
We summarise these results in Table \ref{tab:statistics}.
Both companion galaxies and quasar host galaxies (for which we had masses) show a slight trend where more massive galaxies and quasars have a higher SFR (Fig. \ref{fig:main_sequence}).
This trend agrees with the main sequence of star formation result from \citealt{Duarte_2017}.

Both associated and non-associated companions scatter similarly in sSFR/SFR vs stellar mass space.
To check whether quasar activity can influence their companions, we checked but found no relation between the [O\,\textsc{ii}] luminosity, SFR, and sSFR of the associated galaxies and quasar bolometric luminosity, quasar SMBH mass, and PD.
This is partially due to the fact that the number of associated galaxies is too small to observe a clear trend.
Similarly, since we know stellar masses of the host galaxy for only few quasars \citep{Bettoni_2015}, we did not find a correlation between the quasar bolometric luminosity and host SFR.
In the inset of Fig. \ref{fig:main_sequence} we show the location of the whole sample of galaxies as a function of specific star formation rate and stellar masses (the star formation main sequence, \citealt{Kennicutt_2012}).
For reference, we plot z$\sim$0 and z$\sim$1 trends from literature \citep{Osborne_2020}.

If the triggering was a result of a merger, then massive SFR would be expected to be present in companions and quasars.
However, the overall results from the [O\,\textsc{ii}] line emission in the whole sample show that the SFR rates of the associated galaxies are modest (median SFR value close to Milky Way type galaxy; \citealt{Kennicutt_1998, Kennicutt_2012}).
Only five cases have SFR value similar to starburst galaxies ($>$10 M$_{\odot}$ yr$^{-1}$).
We did not observe any correlation between the SFR of the companion and the quasar [O\,\textsc{ii}] luminosity.

\subsection{Spectra of the host galaxies}

For some QSO we obtained spectra slightly offset from the nucleus.
Considering both slit orientation and brightness of the nucleus, for 8 QSO we were able to gather the light from the host galaxy (see Fig. \ref{fig:host_spectra} in the Appendix \ref{sec:appendix_spectra}).
The H, K and G-band absorption lines are clearly visible.
Additionally, in three cases, i.e. QSO 390, 398 and 403, a clear spectrum of a post-starburst galaxy \citep{Cales_2015} is visible.
All, except QSO 390, exhibit the [O\,\textsc{ii}] emission line.
Note that for five out of eight host detections there is an associated companion galaxy to the QSO.
Together with data in Paper \citetalias{Bettoni_2017} we were able to detect for 11 QSO (8 in this paper) spectral signatures of their host galaxies with 3 cases showing a typical post-starburst spectrum.

\section{Summary and Conclusions}

Combining our new results with those from Paper \citetalias{Bettoni_2017}, we have investigated the close environments of 34 QSO by obtaining optical spectroscopy of the companion objects in the QSO fields to probe for physical association and to assess the recent star formation events.
For about half (15 out of 34) of the observed QSO we found at least one associated companion galaxy.
The associated galaxies showed modest SFR (median SFR value similar to a Milky Way type galaxy) based on [O\,\textsc{ii}] emission line.
For five (out of 18) associated galaxies, the SFR level was that of a starburst.
We were able to obtain the host galaxy spectra for 11 QSO, with 3 cases showing post-starburst signatures.

All together these data confirm the finding of our previous Paper \citetalias{Bettoni_2017} that the role of the QSO activity in the SFR of the associated companion is very modest.
Further work in this field should include comparison of quasar environments to the environments of a control group of low-redshift inactive galaxies of similar luminosity as the host of QSO.

\section*{Acknowledgements}
We thank the anonymous Referee for their careful review and for providing comments which improved the quality of this manuscript.
MBS and JK acknowledge financial support from the Academy of Finland, grant 311438.
MBS also acknowledges financial support from Finnish Centre for Astronomy with ESO (FINCA) and Academy of Finland grant 320045.
MBS thanks Kari Nilsson for helpful discussions and software on determining the significance of spectral lines and Roberto de Propris and Mauri Valtonen for literature suggestions.
Two of the target fields were observed during the Nordic Optical Telescope Observing School in October 2017 as part of the University of Turku Astronomy course.

The data presented here were obtained [in part] with ALFOSC, which is provided by the Instituto de Astrofisica de Andalucia (IAA) under a joint agreement with the University of Copenhagen and NOTSA.
Based on observations made with the Nordic Optical Telescope, operated by the Nordic Optical Telescope Scientific Association at the Observatorio del Roque de los Muchachos, La Palma, Spain, of the Instituto de Astrofisica de Canarias.
Based on observations made with the Gran Telescopio Canarias (GTC), instaled in the Spanish Observatorio del Roque de los Muchachos of the Instituto de Astrofísica de Canarias, in the island of La Palma.

Funding for the Sloan Digital Sky Survey IV has been provided by the Alfred P. Sloan Foundation, the U.S. Department of Energy Office of Science and the Participating Institutions. SDSS-IV acknowledges support and resources from the center for High-Performance Computing at the University of Utah. The SDSS web site is www.sdss.org.

SDSS-IV is managed by the Astrophysical Research Consortium for the 
Participating Institutions of the SDSS Collaboration including the 
Brazilian Participation Group, the Carnegie Institution for Science, 
Carnegie Mellon University, the Chilean Participation Group, the French Participation Group, Harvard-Smithsonian center for Astrophysics, 
Instituto de Astrof\'isica de Canarias, The Johns Hopkins University, Kavli Institute for the Physics and Mathematics of the Universe (IPMU) / 
University of Tokyo, the Korean Participation Group, Lawrence Berkeley National Laboratory, 
Leibniz Institut f\"ur Astrophysik Potsdam (AIP), 
Max-Planck-Institut f\"ur Astronomie (MPIA Heidelberg), 
Max-Planck-Institut f\"ur Astrophysik (MPA Garching), 
Max-Planck-Institut f\"ur Extraterrestrische Physik (MPE), 
National Astronomical Observatories of China, New Mexico State University, 
New York University, University of Notre Dame, 
Observat\'ario Nacional / MCTI, The Ohio State University, 
Pennsylvania State University, Shanghai Astronomical Observatory, 
United Kingdom Participation Group,
Universidad Nacional Aut\'onoma de M\'exico, University of Arizona, 
University of Colorado Boulder, University of Oxford, University of Portsmouth, 
University of Utah, University of Virginia, University of Washington, University of Wisconsin, 
Vanderbilt University and Yale University.

\section*{Data Availability}

The data underlying this article which was collected with the NOT are available through the NOT Archive at \url{http://www.not.iac.es/archive/}, and can be accessed with Proposal IDs 56-018 and 57-002. 
The data underlying this article which was collected with the GTC will be shared on reasonable request to the corresponding author.

\bibliographystyle{mnras}
\bibliography{stone_2020.bib} 

\appendix
\section{Notes to individual objects}
\label{sec:fields}

\textbf{QSO 37} \\
Our slit intercepted two possible companions of QSO 37. 
Object B (z=0.3862) shows H \textbeta{} and [O\,\textsc{iii}] emission lines.
The G-band absorption line and [O\,\textsc{ii}] emission line are clearly visible as well.
Gal. B has a SFR typical of luminous infrared galaxy (LIRG).
Object A turned out to be a star.

\textbf{QSO 64} \\
This radio-quiet quasar is at z = 0.3737.
Its SFR is highest compared to other QSO in this work (see \citealt{Kalfountzou_2012}).
Both Gal. A and B are background galaxies at z=0.462 and z=0.5016 respectively.
The slit also intercepted another foreground galaxy at a PD=1.5 Mpc, at redshift z=0.1268.
The slit also intercepted three stars, which were excluded from further analysis.

\textbf{QSO 95} \\
Close to this QSO (at redshift 0.2272) there are some early-type galaxies; from SDSS we can see that they form a group at z$\sim$0.12 not connected with the QSO.
Galaxy A is a background object. 

\textbf{QSO 165} \\
The QSO is at redshift 0.3965.
Gal. A (z=0.3957) exhibits [O\,\textsc{ii}], H \textbeta{} and [O\,\textsc{iii}] emission as well as Ca\,\textsc{ii} absorption lines.
Gal. A is a true companion galaxy for this quasar with SFR typical of LIRG.
For this QSO we detect the host galaxy spectrum shown in Fig. \ref{fig:host_spectra}.

\textbf{QSO 199} \\
This QSO (z=0.2510) is hosted by a face-on spiral galaxy.
The closest galaxy, Gal. A (PD=231 kpc), is a foreground emission line galaxy at z=0.069.

\textbf{QSO 239} \\
In the immediate environment of this QSO (z=0.3495) there are 5 galaxies.
We observed the spectrum of objects labelled A, B and D (PD=69, 96 and 112 kpc respectively).
Galaxy A is physically associated to the QSO, while B and D are foreground galaxies.
Gal. A spectrum is noisy and the [O\,\textsc{ii}] emission line EW is below the 3-sigma of the noise level.

\textbf{QSO 304} \\
The radio-quiet QSO is at z=0.3857.
None of the observed possible companions turned out to be associated with the QSO.
Gal. A (z=0.2837) and Gal. B (z=0.2821) have very similar redshift and are a probable foreground pair.
Gal. C (z=0.1697) is a foreground galaxy.
Gal. E is a background object.

\textbf{QSO 329} \\
The QSO is at redshift 0.3844 with visible strong [O\,\textsc{iii}] and [O\,\textsc{ii}] emission lines, K and H absorption lines as well as the G-band.
Gal. A (z=0.3912) exhibits prominent K, H, G-band, H \textbeta{} absorption lines and [O\,\textsc{ii}] emission line (SFR typical of LIRG).
Gal. B (z=0.3947) exhibits also K, H, G-band, H \textbeta{} absorption lines. Both are background galaxies.

\textbf{QSO 362} \\
The QSO (z=0.4277) exhibits emission lines and Ca\,\textsc{ii} absorption doublet.
Gal. A (z=0.2777) is a foreground starburst galaxy.

\textbf{QSO 364} \\
The QSO (z=0.3461) exhibits emission lines and Ca\,\textsc{ii} absorption doublet.
Gal. A (z=0.3467) is a companion galaxy with a prominent [O\,\textsc{ii}] emission line and Ca\,\textsc{ii} absorption lines.
Its SFR is typical of ultraLIRGs.

\textbf{QSO 365} \\
The QSO (z=0.4578) also has a companion starburst galaxy (Gal. A) at z=0.4567 with a prominent [O\,\textsc{ii}] emission line (SFR typical of a LIRG).
Object B (z=0.1078) is a foreground galaxy.

\textbf{QSO 373} \\
Gal. A (z=0.1868) of this QSO (z=0.4543) is a foreground galaxy.

\textbf{QSO 376} \\
Gal. A (z=0.4106) and Gal. B (0.4105) around QSO 376 (z=0.3491) are both background galaxies and probably form a galaxy pair since their redshifts are similar.

\textbf{QSO 387} \\
Gal. A (z=0.3015) and Gal. C (z=0.3004) around QSO 387 (z=0.3034) are not associated with the quasar and are both dominated by strong absorption lines in their spectra. A foreground galaxy Gal. Y (z=0.0919) was intercepted by slit 2. Note that Gal. Y falls outside of Fig. \ref{fig:ima} due to the large angular separation from QSO.

\textbf{QSO 390} \\
The QSO (z=0.2369) exhibits prominent K, H, Balmer, G-band and Mgb absorption lines.
Gal. A is a companion galaxy at redshift 0.2370 and it exhibits absorption as well as emission lines.
Its [O\,\textsc{ii}] emission line EW is below 3-sigma level of the noise.

\textbf{QSO 394} \\
Gal. A (z=0.3135) is a companion galaxy of QSO 394 (z=0.3123) with an [O\,\textsc{ii}] emission line and several absorption lines.
However, its SFR is modest, typical of a red galaxy.
Gal. C (z=0.0888) is a foreground galaxy.

\textbf{QSO 398} \\
Gal. A (z=0.4381?) has a noisy spectrum, identified as a background galaxy of QSO 398 (z=0.4297) based on Ca\,\textsc{ii} absorption lines.

\textbf{QSO 401} \\
Gal. A (z=0.3750) and Gal. B (z=0.3389) are both foreground galaxies of QSO (z=0.3960).

\textbf{QSO 403} \\
Gal. A (z=0.3076) is a foreground galaxy to QSO 403 (z=0.4370), dominated by prominent absorption lines.

\textbf{QSO 404} \\
Gal. A (z=0.4263) is a foreground starburst galaxy to QSO 404 (z=0.4601), showing a strong Ca\,\textsc{ii} absorption lines.
Note that Gal. A of QSO 404 is at PD=574 kpc, which is $>$ 250 kpc, our selection criterion. Our possible companion was a very faint object close to the QSO, for which we cannot obtain a good spectrum. 
Since the spectra of Gal. A with larger PD was obtained during the same exposure we included it in this work.

\textbf{QSO 406} \\
The very faint Gal. A ($m_r$=21.22) turned out to be a foreground object at z=0.2486.

\textbf{QSO 409} \\
Gal. A (z=0.4678 ?) is a background galaxy to QSO 409 (z=0.3766).

\section{Optical spectra}
\label{sec:appendix_spectra}
\begin{figure*}
 \includegraphics[width=17cm]{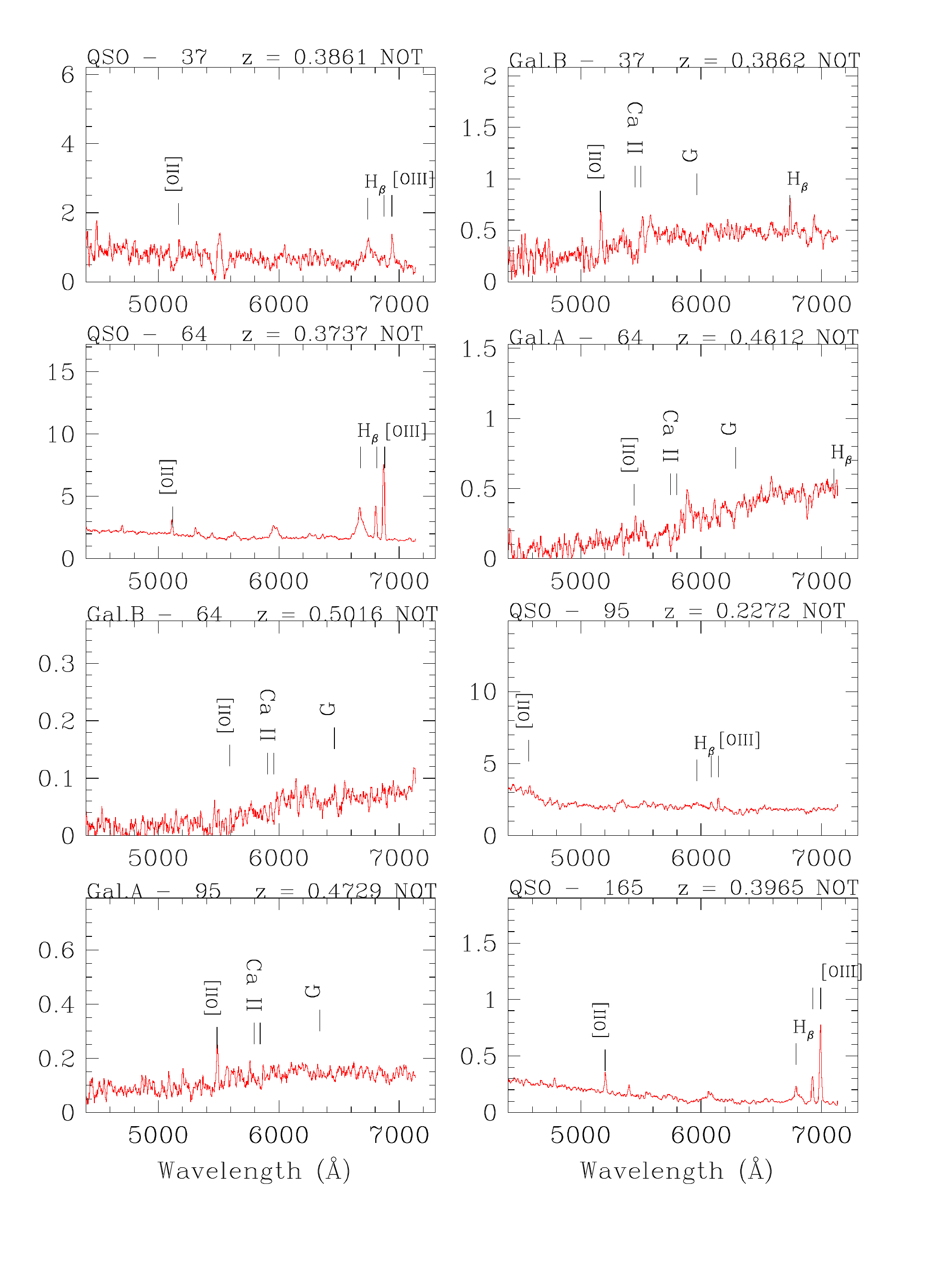}
\caption{Optical spectra of quasars and companion galaxies (smoothed), showing wavelength ({\AA}) versus flux (10$^{-16}$ erg cm$^{-2}$ s$^{-1}$ {\AA}$^{-1}$).}
\label{fig:spectra}
\end{figure*}

The final spectra are given in Fig. \ref{fig:spectra}.
The off-nucleus spectra of QSO hosts are presented in Fig. \ref{fig:host_spectra}.

\renewcommand{\thefigure}{B\arabic{figure} (Cont.)}
\addtocounter{figure}{-1}

\begin{figure*}
 \includegraphics[width=17cm]{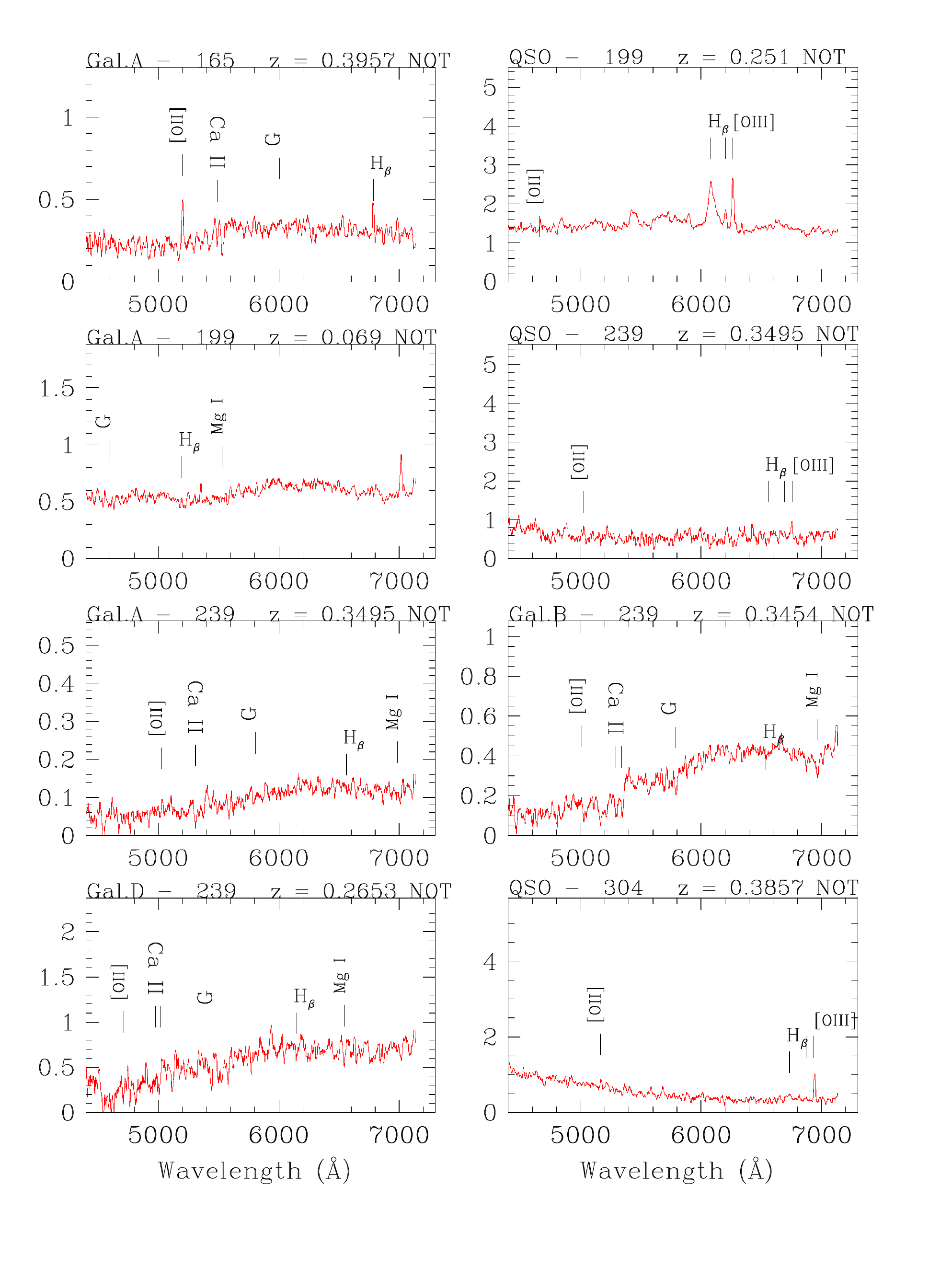}
\caption{}
\end{figure*}

\renewcommand{\thefigure}{B\arabic{figure} (Cont.)}
\addtocounter{figure}{-1}

\begin{figure*}
 \includegraphics[width=17cm]{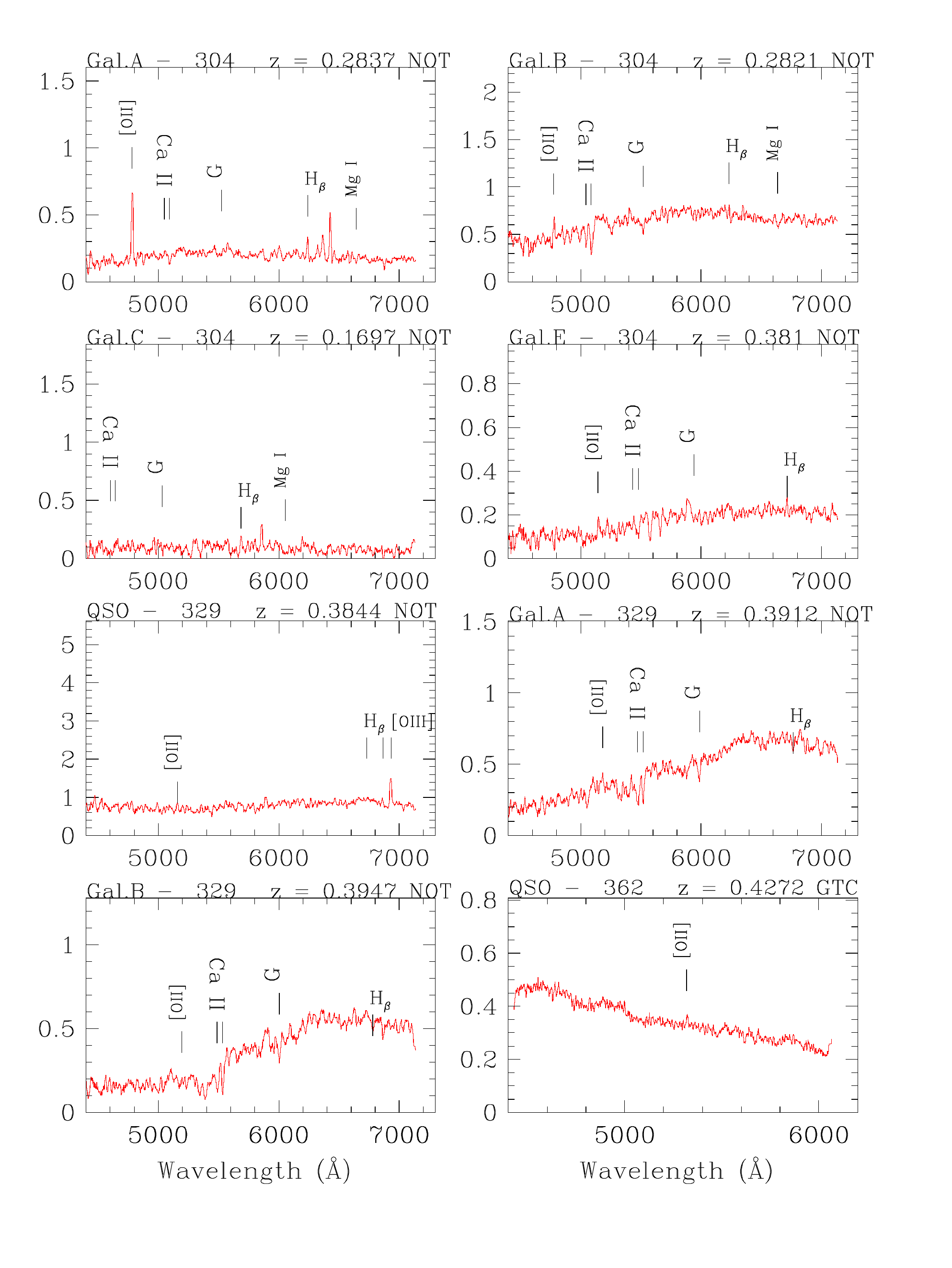}
\caption{}
\end{figure*}

\renewcommand{\thefigure}{B\arabic{figure} (Cont.)}
\addtocounter{figure}{-1}

\begin{figure*}
 \includegraphics[width=17cm]{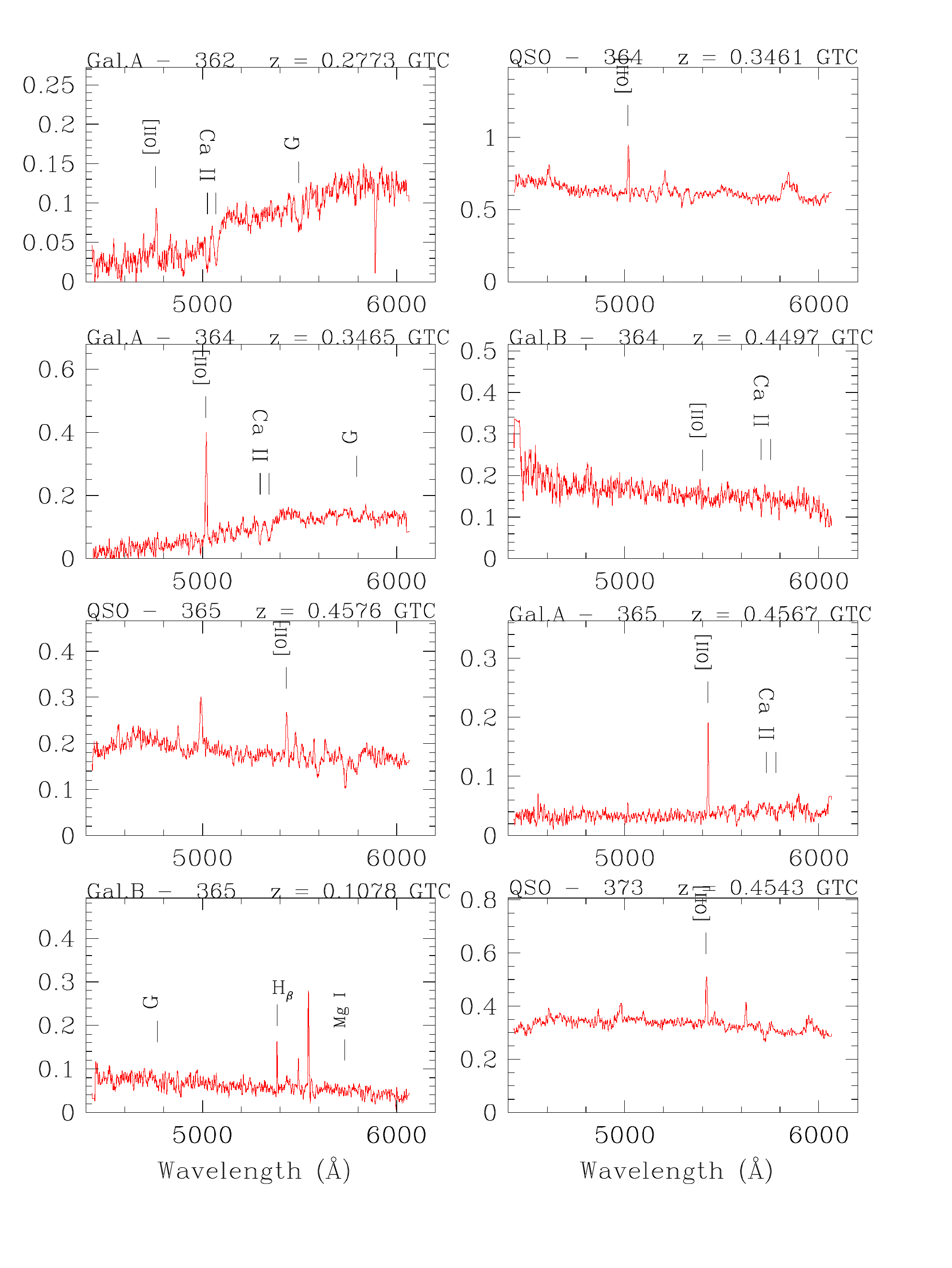}
\caption{}
\end{figure*}

\renewcommand{\thefigure}{B\arabic{figure} (Cont.)}
\addtocounter{figure}{-1}

\begin{figure*}
 \includegraphics[width=17cm]{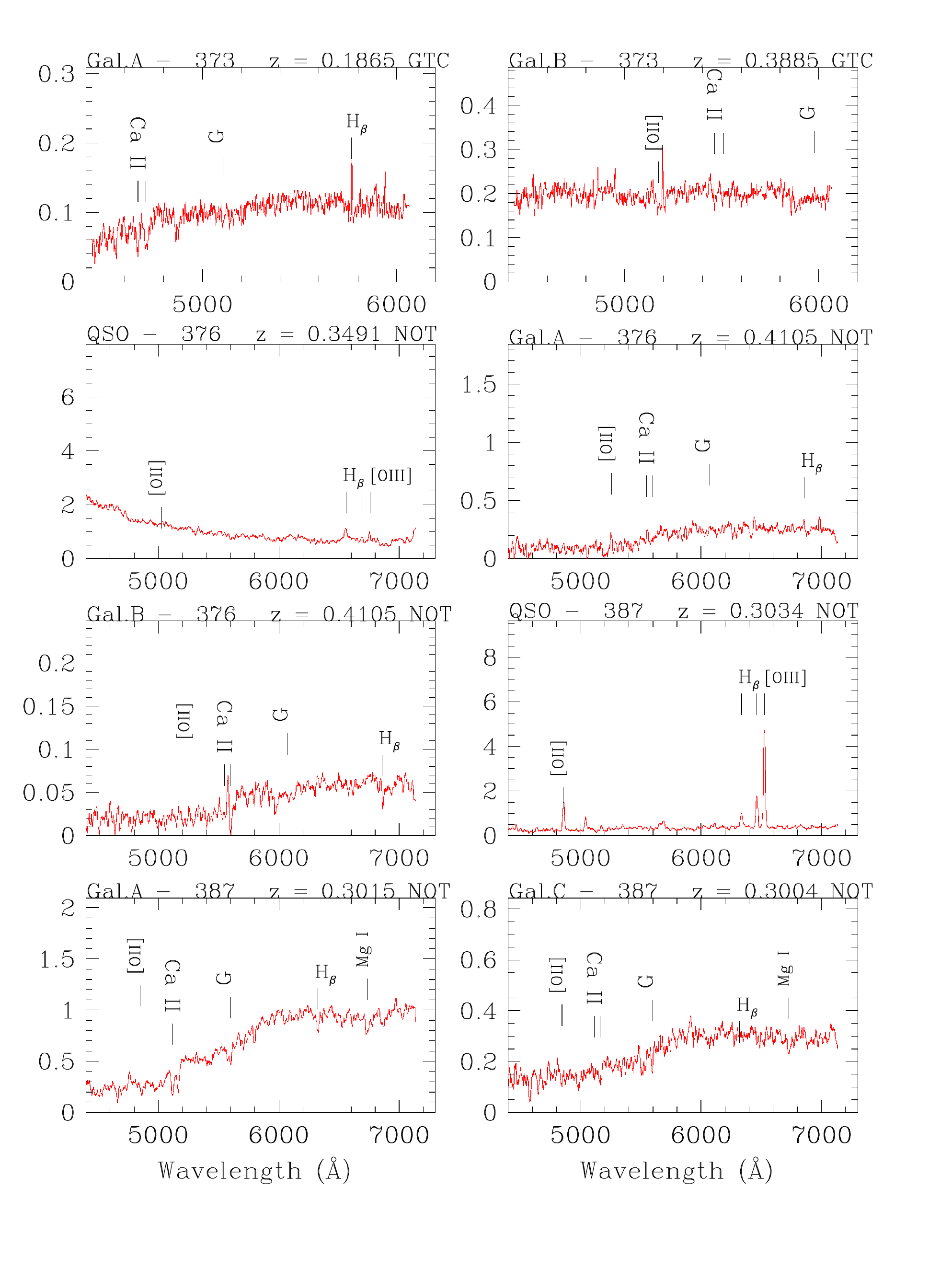}
\caption{}
\end{figure*}

\renewcommand{\thefigure}{B\arabic{figure} (Cont.)}
\addtocounter{figure}{-1}

\begin{figure*}
 \includegraphics[width=17cm]{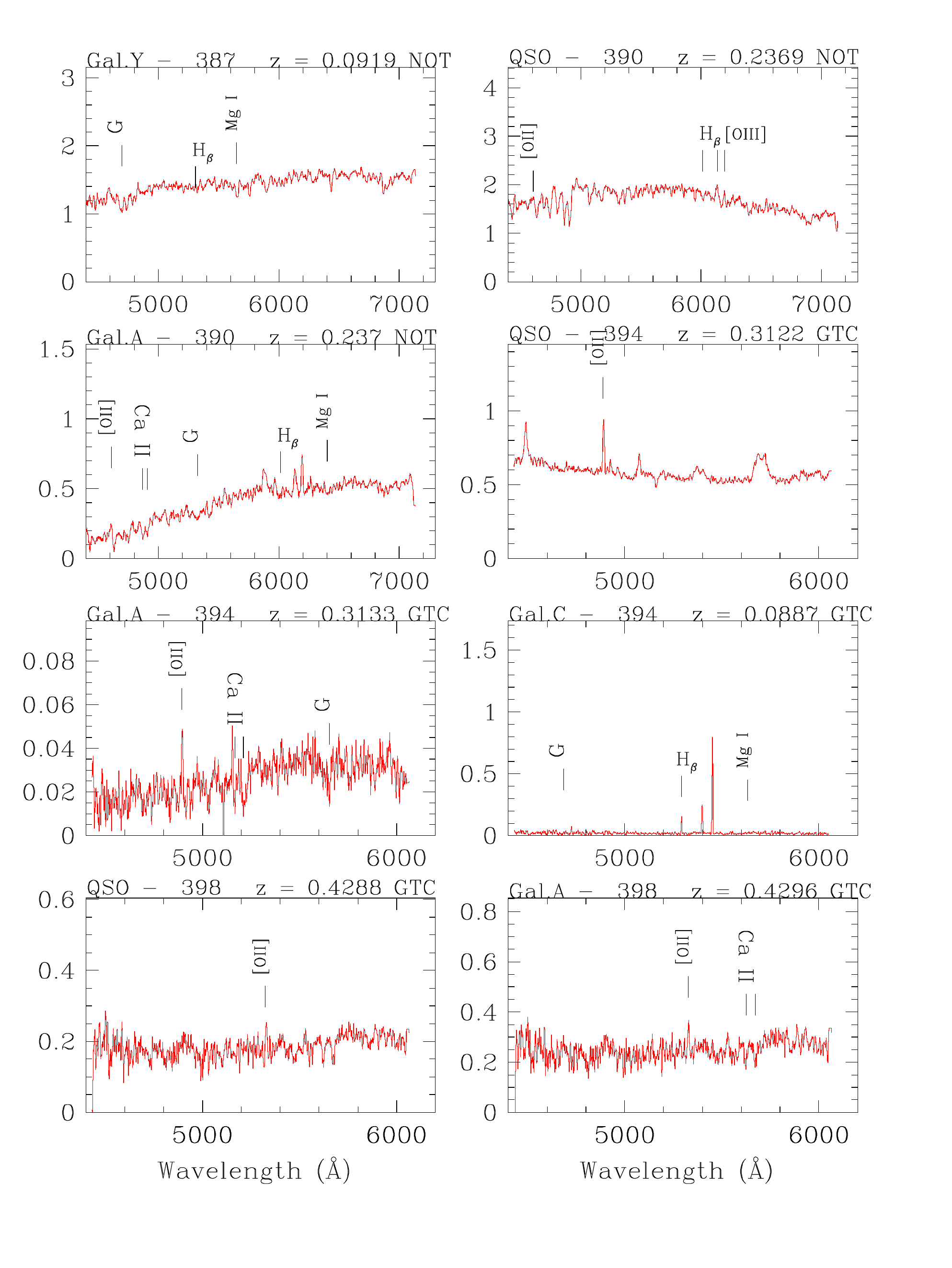}
\caption{}
\end{figure*}

\renewcommand{\thefigure}{B\arabic{figure} (Cont.)}
\addtocounter{figure}{-1}

\begin{figure*}
 \includegraphics[width=17cm]{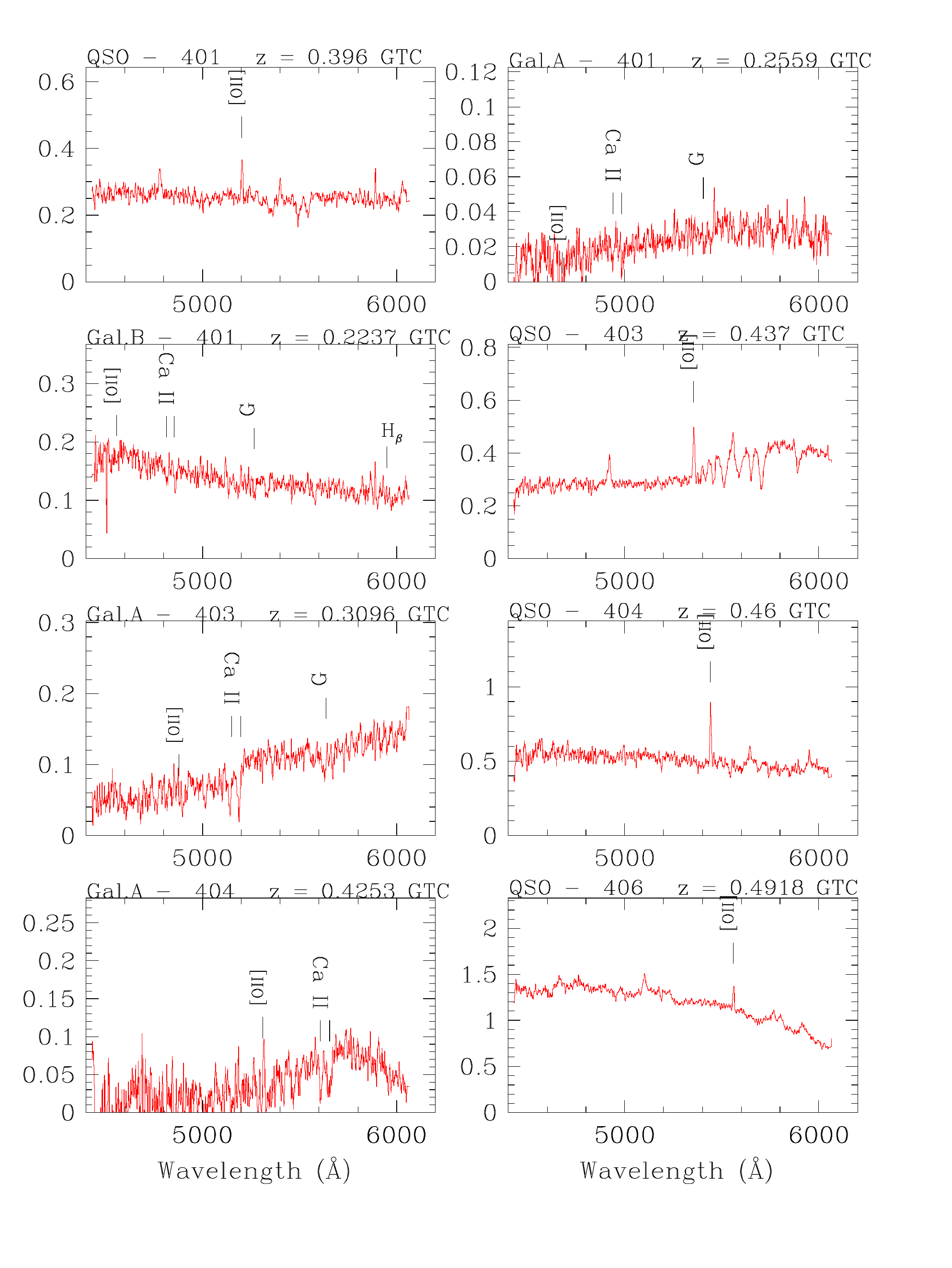}
\caption{}
\end{figure*}

\renewcommand{\thefigure}{B\arabic{figure} (Cont.)}
\addtocounter{figure}{-1}

\begin{figure*}
 \includegraphics[width=17cm]{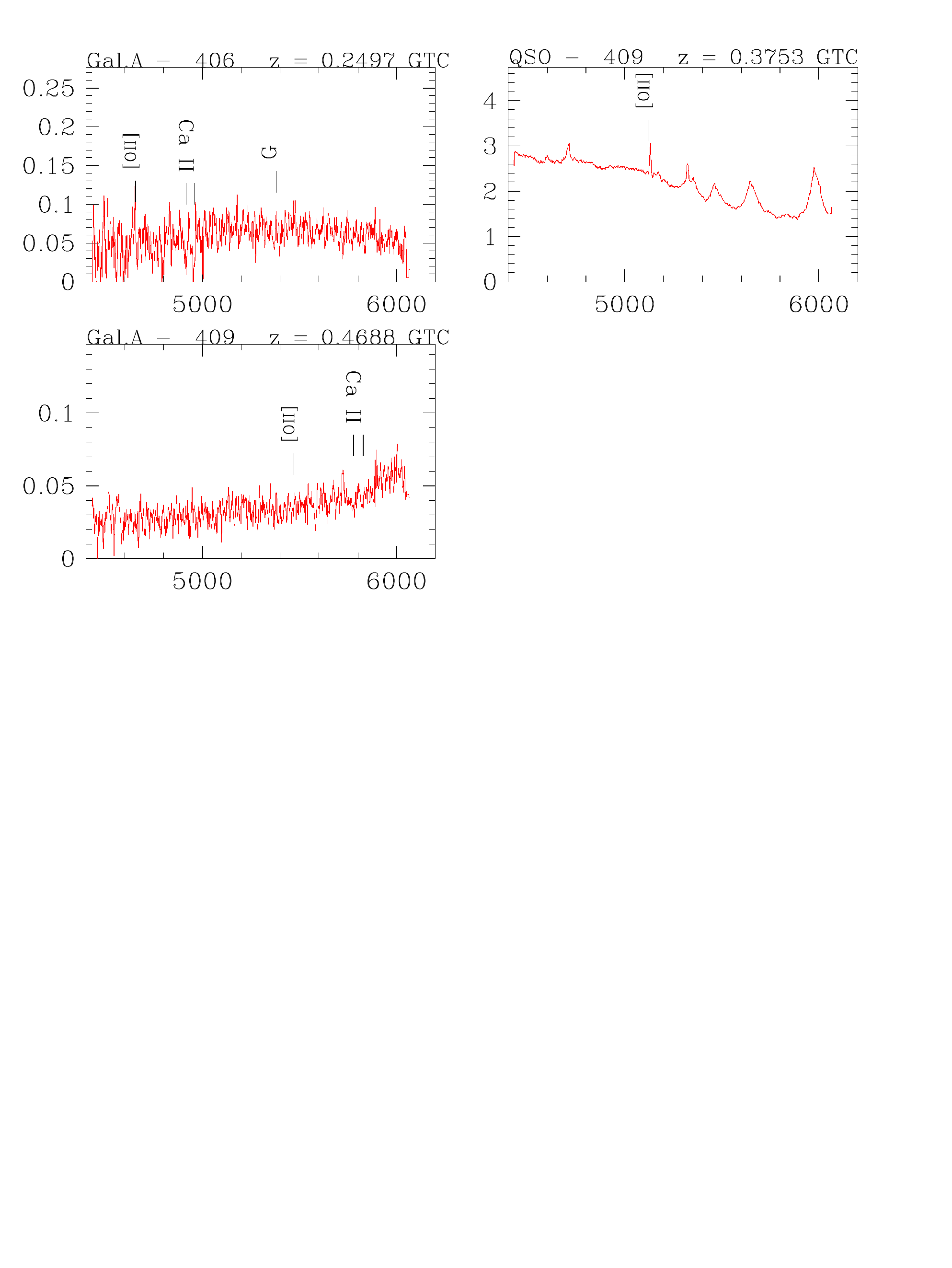}
\caption{}
\end{figure*}

\renewcommand{\thefigure}{B\arabic{figure}}

\begin{figure*}
 \includegraphics[width=15.5cm]{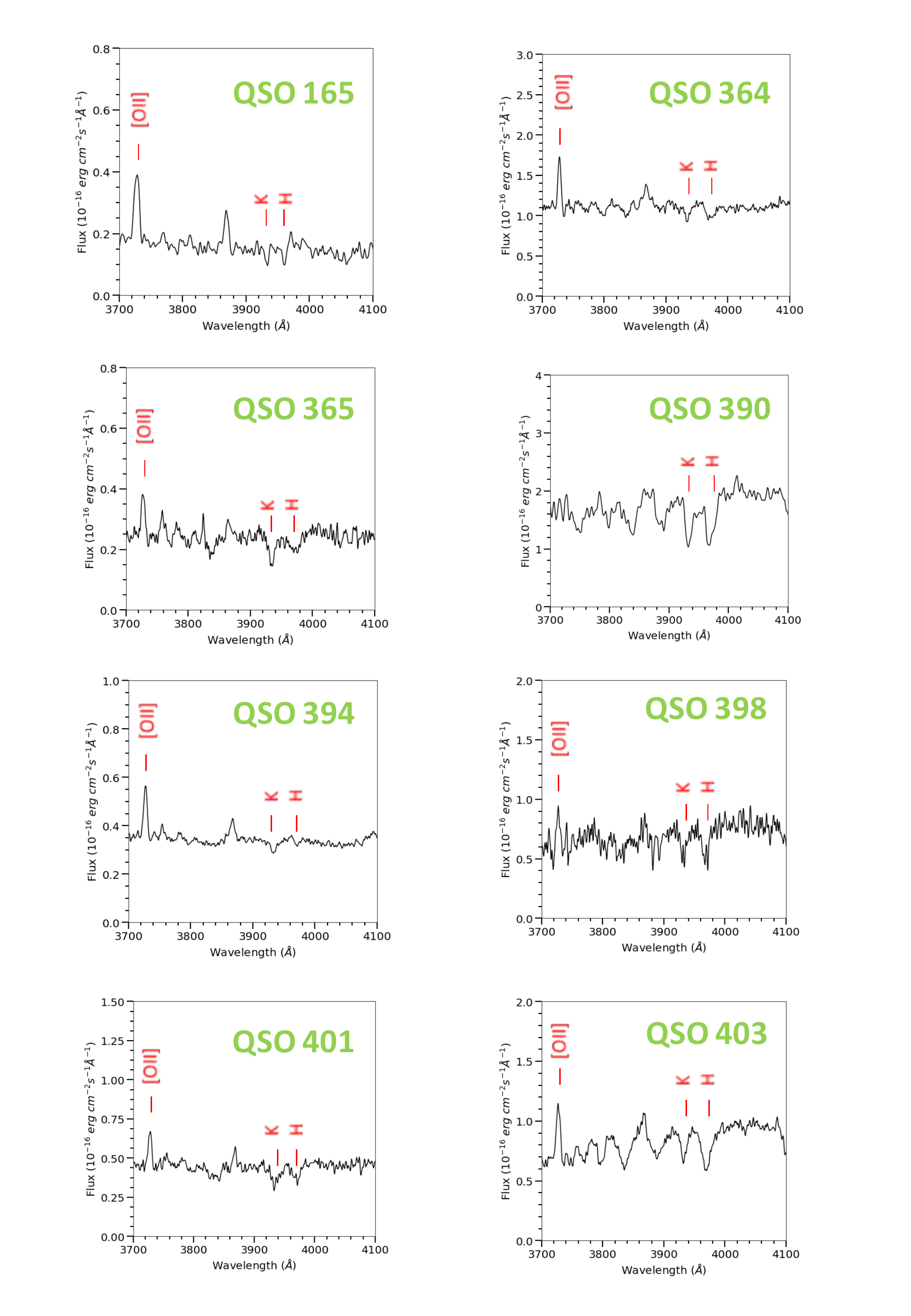}
\caption{Rest-frame spectra of the QSO host galaxies.}
\label{fig:host_spectra}
\end{figure*}


\bsp	
\label{lastpage}
\end{document}